%% file: H2394.tex
\documentclass{aa}

\usepackage{epsf}
\usepackage{graphics}
\usepackage{lscape}

\begin{document}

\input clipfig.tex
\useunitmm

\def \charthoffset  {\hspace{0.2cm}}
\def \charthsep     {\hspace{0.3cm}}
\def \chartvsep     {\vspace{0.3cm}}
\newcommand{\putcharta}[1]{\clipfig{#1}{87}{0}{52}{199}{245}}
\newcommand{\chartlinea}[2]{\parbox[t]{18cm}{

\noindent\charthoffset\putcharta{#1}
\charthsep\charthsep\putcharta{#2}}}


\newcommand{\lb}{$\lambda$}
\newcommand{\sm}[1]{\footnotesize {#1}}
\newcommand{\inft}{$\infty$}
\newcommand{\vlv}{$\nu L_{\rm V}$}
\newcommand{\lv}{$L_{\rm V}$}
\newcommand{\lx}{$L_{\rm x}$}
\newcommand{\lsoft}{$L_{\rm 250eV}$}
\newcommand{\lhard}{$L_{\rm 1keV}$}
\newcommand{\vlsoft}{$\nu L_{\rm 250eV}$}
\newcommand{\vlhard}{$\nu L_{\rm 1keV}$}
\newcommand{\vlir}{$\nu L_{60\mu}$}
\newcommand{\ax}{$\alpha_{\rm x}$}
\newcommand{\aopt}{$\alpha_{\rm opt}$}
\newcommand{\aoxs}{$\alpha_{\rm oxs}$}
\newcommand{\aoxh}{$\alpha_{\rm oxh}$}
\newcommand{\airhard}{$\alpha_{\rm 60\mu-hard}$}
\newcommand{\aoxsoft}{$\alpha_{\rm ox-soft}$}
\newcommand{\aio}{$\alpha_{\rm io}$}
\newcommand{\aixs}{$\alpha_{\rm ixs}$}
\newcommand{\aixh}{$\alpha_{\rm ixh}$}
\newcommand{\hb}{H$\beta$}
\newcommand{\nh}{$N_{\rm H}$}
\newcommand{\nhgal}{$N_{\rm H,gal}$}
\newcommand{\nhfit}{$N_{\rm H,fit}$}
\newcommand{\ale}{$\alpha_{\rm E}$}
\newcommand{\cts}{$\rm {cts\,s}^{-1}$}
\newcommand{\pl}{$\pm$}
\newcommand{\kev}{\rm keV}
\newcommand{\rb}[1]{\raisebox{1.5ex}[-1.5ex]{#1}}
\newcommand{\ten}[2]{#1\cdot 10^{#2}}
\newcommand{\msun}{$M_{\odot}$}
\newcommand{\dM}{\dot M}
\newcommand{\dMM}{$\dot{M}/M$}
\newcommand{\dMedd}{\dot M_{\rm Edd}}
\newcommand{\kms}{km\,$\rm s^{-1}$}

\thesaurus{03(02.01.2; 11.01.2; 11.14.1; 11.19.1; 11.17.3)}

\title{X-ray variability in a complete sample of Soft X-ray selected AGN
\thanks{Based in part on
observations at the European Southern Observatory La Silla (Chile) with
the 2.2m telescope of the Max-Planck-Society during MPG and ESO time
 and the ESO 1.52m telescope.}
}
\author{D. Grupe\inst{1, }\thanks{
Guest Observer, McDonald Observatory,
University of Texas at Austin},
H.-C. Thomas\inst{2},
\and K. Beuermann\inst{1,3},
}
\offprints{\\ D. Grupe (dgrupe@xray.mpe.mpg.de)}
\institute{
MPI f\"ur extraterrestrische Physik, Postfach 1312, 85741 Garching, FRG
\and MPI f\"ur Astrophysik, Karl-Schwarzschild-Str. 1, 85741 Garching, FRG
\and Universit\"ats-Sternwarte, Geismarlandstr. 11, 37083 G\"ottingen,
FRG
}
\date{Received 11 August 2000; Accepted 04 December 2000}
\maketitle
\markboth{D. Grupe et al.: X-ray variability of Soft X-ray AGN}{ }

\begin{abstract}
We present ROSAT All-Sky Survey and ROSAT pointed observations (PSPC and HRI)
of a complete sample 
of 113 bright soft X-ray AGN selected from the ROSAT Bright Source Catalog.   
We compare these
observations in order to search for extreme cases of flux and spectral X-ray
variability - X-ray transient AGN. Three definite transients and one
transient candidate are found.
 The
other sources show amplitude variations typically by factors of 2-3 on
timescales of years. We found that the variability strength 
on timescales of days is a function of the steepness of the X-ray spectrum:
steeper X-ray objects show stronger variability than flat X-ray spectrum sources. 
We also present new HRI measurements of our extreme X-ray transients IC
3599 and WPVS007. We discuss possible models to explain the X-ray
transience and the variabilities observed in the non-transient sources.
\end{abstract}
\section{Introduction}

The optical-UV-soft X-ray bump has turned out to be a common property
of most Narrow-Line Seyfert 1 (NLS1)
type Active Galactic Nuclei (AGN).  With ROSAT's
(Tr\"umper 1983) Position Sensitive Proportional Counter (PSPC, Pfeffermann et
al. 1986) with its spectral sensitivity to energies below 0.5 keV,
numerous AGN have been found that show a soft X-ray
excess, commonly believed to be the high energy part of the ``Big Blue/UV 
Bump''. This bump emission is thought to be produced by
 an accretion disk that surrounds the central black hole. The soft X-ray
 emission is explained by Compton scattering of thermal UV disk
 photons by a hot electron layer above the disk. The UV photon spectrum is
 shifted further into the soft X-ray range as the engine accretes closer
 to its Eddington accretion rate (e.g. Ross et al. 1992). 
Before ROSAT, a study of soft X-ray selected AGN had to rely on
serendipitous observations, notably with the EINSTEIN Image
Proportional Counter (C\'ordova et al. (1992), Puchnarewicz et al. (1992))

The ROSAT All-Sky Survey 
(RASS, Voges et al. 1999) and later re-observations of many sources made it possible
to study the long-term behaviour of AGN in X-rays. 
AGN can vary in three ways, either by changes in the
flux (or count rate), solely by spectral changes, or a combination of
both. Most common are 
variations in X-ray flux by factors of 
2-3 on timescales of days and years (e.g. Lawrence et al. 1977, Green et
al. 1993, Boller et al. 1996). 
However,  factors of more than ten have
been reported for several sources (e.g. IRAS 13224--3809, 
Boller et al. 1997 or PHL 1092, Brandt et al. 1999). AGN can also change the
shape of their X-ray spectra. The most extreme example of such a
spectral change has been reported on the Narrow-line Seyfert 1 galaxy (NLS1) RX
J0134.2--4258 (Grupe et al. 2000a, Komossa \& Meerschweinchen 2000). 
 It has been
shown that the timescales of the variabilities found in AGN 
scale with luminosity (Barr \& Mushotzky 1986, Lawrence \& Papadakis 1993, Green
et al. 1993).
Leighly (1999a,b) has presented a comprehensive variability study on a sample of
25 AGN observed by ASCA and found a) that the variability strength is a function
of the luminosity and b) that NLS1 are more variable than Broad-Line Seyfert 1s.
Boller et al. (1996) found in their NLS1 study based on ROSAT data that the
timescale of the variability is a function of the luminosity. 
Variability in X-rays can be due either 
to changes in the absorption column or intrinsic variability of the X-ray
source. Absorption by both neutral and ionized gas 
has been proposed to explain
the X-ray variability observed (see Abrassart \& Czerny 2000 and Komossa
\& Meerschweinchen 2000). Intrinsic variability can be caused by e.g. changes in
the accretion rate or relativistic beaming effects (e.g.
Boller et al. 1997). 

With the RASS a new AGN phenomenon has
been established: X-ray transience.
X-ray transience is the most extreme form of variability in AGN. 
On timescales of years, the count rates decreased by factors of more than 100
or the sources even vanished in X-rays.
 The first example of an X-ray transient AGN was given by
Piro et al. (1988) who reported a change in the soft X-ray luminosity of the
Seyfert 1 galaxy E1615+061 by two orders of magnitude on timescales of years.
Transient sources are thought to accrete at high rates, close to the 
Eddington limit, and therefore have very soft X-ray spectra.
The RASS with its sensitivity to energies down to 0.1 keV
has a high potential to find 
transient sources. Transience in AGN can be caused by
changes of accretion disk properties (e.g. the temperature, see WPVS007, Grupe
et al. 1995b), a dramatic increase of the accretion rate (e.g. in an outburst as
seen in IC 3599, Brandt et al. 1995, Grupe et al. 1995a), or alternatively,
changes in the absorption column in the line of sight.

Thomas et al. (1998) have presented a completely identified 
sample of all 397 bright soft X-ray
selected, high galactic latitude RASS sources ( mean PSPC count rate 
$\rm \ge 0.5~cts~s^{-1}$, hardness ratio 1 
(HR1) $<$ 0.00\footnote{The hardness ratio is defined as HR1 = 
(hard-soft)/(hard+soft) with
soft PSPC channels = 11-41 and hard = 52-201.}, and 
$\rm |b|~>~20^{\circ}$, based on RASS II). Of these sources, 
113 are
AGN, which is the sample we present here. Our sample is complete for all AGN
following the criteria above.
BL Lac objects were excluded  because of their different X-ray emission
mechanism. 

The task of this new paper is to compare the
RASS results with measurements of ROSAT pointed observations in order to find 
new X-ray transient AGN.
In Sect. \ref{obs} we describe the observations and data reduction. Sect.
\ref{results} shows the 
 results obtained for the whole sample and 
lists individual X-ray transient AGN, and in Sect. \ref{discuss} we discuss 
the results.
Throughout the paper, luminosities are calculated assuming a Hubble
constant of \mbox{$H_0 = 75$\,km\,s$^{-1}$Mpc$^{-1}$} and a deceleration
parameter of $q_0 = 0$. Spectral slopes, $\alpha$, 
are defined by $F\propto \nu^{-\alpha}$.

\section{\label{obs} Observations and data reduction}

All objects have been observed (by definition) during the RASS. 
The data from RASS II\footnote{The difference between RASS I and RASS II 
is in the
detection likelihood for acceptance of a source;  Voges et al. 1999. 
Meanwhile the RASS III
has been released.)} 
were extracted from the event files
as described in Grupe et al. (1998a).
Pointed PSPC and High Resolution Imager (HRI) 
observations were derived from the ROSAT public data
archive at MPE
Garching (ftp.xray.mpe.mpg.de). The source PSPC count rates were determined
using a weighted exposure map in four energy channel ranges, 8-41, 42-51, 52-90, and
91-201\footnote{One energy channel transfers to 10 eV}. 
Power-law spectral models were applied to the RASS as well as to the pointed
PSPC data. We limit the spectral analysis to single power-law fits only because
of
the limited signal-to-noise ratios of the RASS observations. The absorption
parameter for cold absorption $N_H$ was either fixed to the Galactic value given
by Dickey \& Lockman (1990), or left as a free parameter (see Table
\ref{spec_list}). 

In cases where 
more than one pointed PSPC (or HRI) observation was available in the
archive, we usually took the longest one in order to get better 
signal-to-noise ratio.
HRI count rates were converted into effective PSPC count rates using
 the W3PIMMS 
program of NASA's Goddard Space Flight Center
(version 2.7, 1999, http://heasarc.gsfc.nasa.gov/Tools/w3pimms.html) based on
the power-law fits to the RASS data with the absorption parameter $N_H$ fixed to
the galactic value, and 
assuming no spectral changes between the RASS and the HRI observations. 
For all X-ray reduction and analysis tasks we used
the EXSAS data analysis package of
the MPE Garching (5th edition, Zimmermann et al. 1998).

\section{\label{results} Results}
Table \ref{CR_list} lists the optical coordinates derived
 from the US Naval Observatory scans (USNO A2.0) and 
summarizes the count rates and hardness ratios of the RASS,
pointed PSPC, and HRI observations. The X-ray positions of the objects are given in
Thomas et al. (1998).
 The RASS count rates
and HR1 given in Table \ref{CR_list} differ slightly from the
values in the ROSAT Bright Source Catalogue (BSC, Voges et al. 1999),
because while the definition of the sample is based on the BSC our 
analysis of the
RASS event files were used for all analysis throughout the paper. 
In two cases (RX J1624.9+7554 and RX J2312.5-3404) the
count rates are below the limit of 0.5 \cts and in one case, IIZW136,
the HR1 $>$ 0.00. 
All observations used for this publication,
with their observing dates, exposure times and the ROSAT Observation Request number (ROR) 
of the pointed observations, are listed in  Table \ref{obs_list}.

\subsection{\label{cr-text} Flux variability}

\begin{figure}
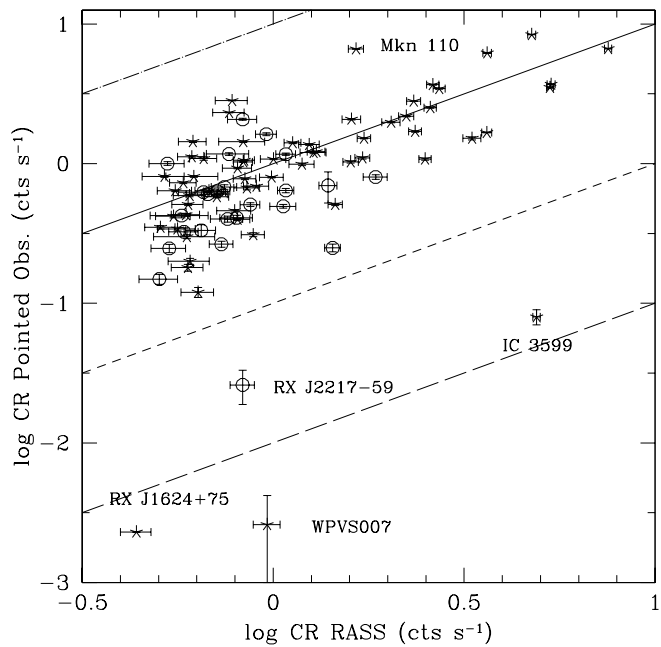

\parbox[h]{8.7cm}{
\clipfig{H2394_F1}{87}{06}{54}{200}{248}
} 
\caption[ ] {\label{cr-cr} RASS vs. pointed observation count rate. PSPC
observations are marked as stars and HRI observations as open circles.
The solid line represents no change between RASS and pointed observation, 
the dot-dashed line represents to a factor of 10,
the short-dashed line to a factor of 0.1, and the long-dashed line to 
a factor of 0.01. The HRI count rates were converted
into PSPC count rates (see section \ref{obs})}
\end{figure}

Fig. \ref{cr-cr} compares the RASS vs. PSPC and HRI pointed observation count 
rates.  In general, if PSPC and HRI
observations were available for one source, preference was given to the PSPC
observation, because RASS and PSPC pointings were performed by the same
type of detector and allow, therefore, a direct comparison without any
assumptions about the spectral shape that is necessary for 
converting the HRI
observations.  The plot displays that the
distribution of the converted HRI count rates does not differ significantly 
from the PSPC data. It shows that the method of conversion
is basically correct.

\begin{landscape}

{\small
\begin{table}
\begin{flushleft}
\caption{\label{CR_list}  Optical positions of the AGN taken from the USNO, 
RASS and pointed observation count rates and hardness ratios. For the HRI
observations expected HRI count rates derived from the RASS observations and 
expected PSPC count rtes are listed.}
\hspace*{-1.5cm}
\begin{tabular}{rlllcccccccc}
\hline\noalign{\smallskip}
& & & & & \multicolumn{2}{c}{RASS} & \multicolumn{2}{c}{PSPC pointed} & \multicolumn{3}{c}{HRI pointed} \\
\rb{No.}  & \rb{$\alpha_{2000}$} & \rb{$\delta_{2000}$} & \rb{Name}  & \rb{z} & 
CR & HR1 & CR & HR1  & expected & CR & converted  \\
& h m s & $^{\circ}~^{'}~^{''}$ & & & $\rm cts~s^{-1}$ & & $\rm cts~s^{-1} $ &  & HRI CR & \cts & CR \\
\noalign{\smallskip}\hline\noalign{\smallskip}
1   & 00 06 19.5 &  +20 12 11 & Mkn 335  & 0.026      & 2.33\pl0.10 & --0.07\pl0.04 & 2.60\pl0.01 & --0.04\pl0.02  & 0.660 &
0.426\pl0.007 & 1.504  \\
2   & 00 25 00.2 & --45 29 34 & ESO 242--G8  & 0.059    & 0.79\pl0.08 & --0.48\pl0.07 & 0.47\pl0.01 & --0.43\pl0.09  & 0.183 &
0.060\pl0.006 & 0.260  \\
3   & 00 39 15.8 & --51 17 02 & WPVS 007    & 0.029     & 0.96\pl0.07 & --0.97\pl0.02 &
0.002\pl0.001 & $^1$  & 0.197 & 0.001 & 0.004  \\
4   & 00 57 20.2 & --22 22 57 & RX J0057.2--2223 & 0.062 & 2.72\pl0.10 & --0.58\pl0.03 & 3.28\pl0.03 & --0.59\pl0.05 & 
 --- & --- & --- \\
5   & 00 58 37.4 & --36 06 05 & QSO 0056--36 & 0.165   & 0.75\pl0.06 & --0.31\pl0.06 & 	---	   & ---  & 0.178 &
0.162\pl0.008 & 0.680  \\
6   & 01 00 27.1 & --51 13 54 & RX J0100.4--5113 & 0.062 & 1.06\pl0.08 & --0.24\pl0.07 & 	---          & ---  & 0.272 &
0.127\pl0.005 & 0.495  \\
7   & 01 05 38.8 & --14 16 14 & RX J0105.6--1416 & 0.070 & 1.56\pl0.07 & --0.17\pl0.04 &  ---        & ---           & ---  & --- &
--- \\
8   & 01 17 30.6 & --38 26 30 & RX J0117.5--3826 & 0.225 & 0.98\pl0.07 & --0.52\pl0.04 & ---  & --- & --- & --- & --- \\
9   & 01 19 35.7 & --28 21 32 & MS 0117--28  & 0.349   & 0.60\pl0.06 & --0.72\pl0.05 & 0.48\pl0.01 & --0.79\pl0.02 &  
--- & --- & --- \\
10  & 01 28 06.7 & --18 48 31 & RX J0128.1--1848 & 0.046 &  0.76\pl0.06 & --0.28\pl0.06 & ---   & --- & --- & --- & --- \\
11  & 01 29 10.7 & --21 41 57 & IRAS F01267--217 & 0.093 & 0.83\pl0.07 & --0.45\pl0.06 &  --- & ---  & 0.192 &
0.482\pl0.011 & 2.076  \\
12  & 01 48 22.3 & --27 58 26 & RX J0148.3--2758 & 0.121 & 2.35\pl0.09 & --0.62\pl0.03 & 1.63\pl0.02 & --0.51\pl0.05 & 
--- & --- & --- \\
13  & 01 52 27.1 & --23 19 54 & RX J0152.4--2319 & 0.113 & 1.08\pl0.06 & --0.52\pl0.04 & 	---  & --- & 0.223 &
0.241\pl0.007 & 1.170  \\
14  & 02 30 05.5 & --08 59 53 & Mkn 1044  & 0.017      & 2.23\pl0.10 & --0.06\pl0.04 & 2.10\pl0.03 & --0.02\pl0.05 & 
--- & --- & --- \\
15  & 02 34 37.8 & --08 47 16 & Mkn 1048  & 0.042      & 1.24\pl0.08 & --0.05\pl0.06 & 1.30\pl0.02 &  +0.04\pl0.05  & 0.355 &
0.106\pl0.008 & 0.371  \\
16  & 03 11 18.8 & --20 46 19 & RX J0311.3--2046 & 0.070 & 0.91\pl0.08 & --0.08\pl0.07   &  ---  & --- &  --- & --- & --- \\
17  & 03 19 48.7 & --26 27 12 & RX J0319.8--2627 & 0.079 & 0.52\pl0.07 & --0.52\pl0.09 & 0.78\pl0.01 & --0.35\pl0.06 &  
--- & --- & --- \\
18  & 03 23 15.8 & --49 31 11 & RX J0323.2--4931 & 0.071 & 1.45\pl0.06 & --0.50\pl0.03 & 0.50\pl0.01 & --0.46\pl0.08  & 0.304 &
0.094\pl0.007 & 0.450  \\
19  & 03 25 02.2 & --41 54 18 & ESO 301--G13     & 0.059 & 0.96\pl0.06 & --0.48\pl0.04 & ---	  & --- & 0.219 &
0.371\pl0.012 & 1.625  \\
20  & 03 33 40.2 & --37 06 55 & VCV0331--37      & 0.064 & 0.68\pl0.08 & --0.40\pl0.09 & ---	  & --- & 0.159 &
0.142\pl0.007 & 0.605 \\
21  & 03 49 07.7 & --47 11 04 & RX J0349.1--4711 & 0.299 & 0.53\pl0.06 & --0.70\pl0.05 & ---	   & --- & 0.095 &
0.044\pl0.005 & 0.247  \\
22  & 03 51 41.7 & --40 28 00 & Fairall 1116     & 0.059 & 1.86\pl0.14 & --0.37\pl0.06 & ---	   & --- & 0.474 &
0.205\pl0.009 & 0.803 \\
23  & 04 05 01.7 & --37 11 15 & CTS 44	       & 0.055 & 3.29\pl0.11 & --0.37\pl0.02 & --- & --- &  ---  & --- & --- \\
24  & 04 12 41.5 & --47 12 46 & RX J0412.7--4712 & 0.132 & 0.77\pl0.10 & --0.52\pl0.09 & ---	   & --- & 0.169 &
0.261\pl0.008 & 1.172  \\
25  & 04 26 00.7 & --57 12 02 & 1H 0419+577      & 0.104 & 3.63\pl0.10 & --0.54\pl0.02 & 6.00\pl0.04 & --0.47\pl0.03  & 0.795 &
0.510\pl0.011 & 2.332  \\
26  & 04 30 40.0 & --53 36 56 & Fairall 303      & 0.040 & 1.18\pl0.09 & --0.50\pl0.06 & 0.99\pl0.02 & --0.39\pl0.09 & 
--- & --- & --- \\
27  & 04 37 28.2 & --47 11 30 & RX J0437.4--4711 & 0.052 & 1.13\pl0.06 & --0.60\pl0.03 & 1.35\pl0.02 & --0.51\pl0.05 & 0.230 &
0.461\pl0.005 & 2.255  \\
28  & 04 39 38.7 & --53 11 31 & RX J0439.6--5311 & 0.243 & 0.83\pl0.12 & --0.81\pl0.06 & 1.44\pl0.02 & --0.79\pl0.02 &  
--- & --- & --- \\
29  & 04 41 22.5 & --27 08 20 & H 0439-272       & 0.084 & 0.62\pl0.10 & --0.09\pl0.12 & 0.77\pl0.01 &  +0.09\pl0.07 &  
--- & --- & --- \\
30  & 06 15 49.6 & --58 26 06 & 1 ES 0614-584    & 0.057 & 0.59\pl0.04 & --0.18\pl0.03  & --- & --- & --- & --- & --- \\
31  & 08 59 02.9 &  +48 46 09 & RX J0859.0+4846  & 0.083 & 0.70\pl0.06 & --0.18\pl0.07  & --- & --- & --- & --- & --- \\
32  & 09 02 33.6 & --07 00 04 & RX J0902.5--0700 & 0.089 & 0.62\pl0.06 & --0.22\pl0.07  & --- & --- &  --- & --- & --- \\
33  & 09 25 13.0 &  +52 17 12 & Mkn 110	       & 0.035 & 1.65\pl0.08 & --0.21\pl0.04 & 6.30\pl0.03 & --0.28\pl0.02  & 0.420 &
1.519\pl0.008 & 5.967  \\
34  & 09 56 52.4 &  +41 15 22 & PG 0953+41       & 0.234 & 0.99\pl0.07 & --0.56\pl0.04 & 0.76\pl0.01 & --0.49\pl0.07  & 
--- & --- & --- \\
35  & 10 05 41.9 &  +43 32 41 & RX J1005.7+4332  & 0.178 & 0.73\pl0.05 & --0.64\pl0.04 & ---	   & --- & 0.143 &
0.052\pl0.004 & 0.265  \\
36  & 10 07 10.2 &  +22 03 02 & RX J1007+2203    & 0.083 & 0.62\pl0.06 & --0.09\pl0.06 & --- & --- &  --- & --- & --- \\
37  & 10 13 03.2 &  +35 51 24 & CBS 126	       & 0.079 & 1.43\pl0.07 & --0.45\pl0.03 & ---	   & --- & 0.316 &
0.055\pl0.005 & 0.249  \\
38  & 10 19 00.5 &  +37 52 41 & HS 1019+37       & 0.135 & 0.81\pl0.06 & --0.08\pl0.05 & 0.37\pl0.01 &  +0.02\pl0.03 & 
--- & --- & --- \\
39  & 10 19 12.6 &  +63 58 03 & Mkn 141	       & 0.042 & 0.50\pl0.06 & --0.45\pl0.06 & ---	   & --- & 0.108 & 
0.032\pl0.004 & 0.149  \\
40  & 10 25 31.3 &  +51 40 35 & Mkn 142	       & 0.045 & 1.73\pl0.07 & --0.62\pl0.03 & 1.44\pl0.02 & --0.56\pl0.05 & 
--- & --- & --- \\
\noalign{\smallskip}\hline
\end{tabular}
\end{flushleft}
\end{table}
}
\clearpage


{\small
\begin{table}
\hspace*{-1.5cm}
\begin{tabular}{rlllcccccccc}
\hline\noalign{\smallskip}
& & & & & \multicolumn{2}{c}{RASS} & \multicolumn{2}{c}{PSPC pointed} & \multicolumn{3}{c}{HRI pointed} \\
\rb{No.}  & \rb{$\alpha_{2000}$} & \rb{$\delta_{2000}$} & \rb{Name}  & \rb{z} & 
CR & HR1 & CR & HR1  & expected & CR & converted  \\
& h m s & $^{\circ}~^{'}~^{''}$ & & & $\rm cts~s^{-1}$ & & $\rm cts~s^{-1} $ &  & HRI CR & \cts & CR \\
\noalign{\smallskip}\hline\noalign{\smallskip}
41  & 10 34 38.6 &  +39 38 28 & RX J1034.6+3938  & 0.044 & 2.62\pl0.10 & --0.74\pl0.02 & 3.53\pl0.03 & --0.71\pl0.04  & --- &
--- & --- \\
42  & 11 17 10.1 &  +65 22 07 & RX J1117.1+6522  & 0.147 & 0.58\pl0.05 & --0.72\pl0.04 & ---	   & --- & 0.108 &
0.060\pl0.007 & 0.325  \\
43  & 11 18 30.4 &  +40 25 55 & PG 1115+407      & 0.154 & 0.56\pl0.06 & --0.46\pl0.06 & 0.32\pl0.01 & --0.51\pl0.10 & 0.121 &
0.068\pl0.005 & 0.315  \\
44  & 11 19 08.7 &  +21 19 18 & Ton 1388         & 0.177 & 1.00\pl0.08 & --0.53\pl0.05 & 1.05\pl0.01 & --0.46\pl0.03  & 0.213 &
0.339\pl0.014 & 1.594 \\ 
45  & 11 31 04.8 &  +68 51 53 & EXO 1128+69      & 0.045 & 1.59\pl0.06 & --0.45\pl0.03 & ---  & --- & --- & --- & --- \\
46  & 11 31 09.5 &  +31 14 06 & B2 1128+31       & 0.289 & 0.56\pl0.06 & --0.11\pl0.08 & ---  & --- & --- & --- & --- \\
47  & 11 38 49.6 &  +57 42 44 & SBS 1136+579     & 0.116 & 0.60\pl0.06 & --0.47\pl0.06 & 0.16\pl0.01 & --0.63\pl0.22 &  
--- & --- & --- \\
48  & 11 39 13.9 &  +33 55 51 & Z 1136+3412      & 0.033 & 0.85\pl0.07 & --0.33\pl0.06 & 0.64\pl0.02 & --0.42\pl0.15 & 
--- & --- & --- \\
49  & 11 41 16.2 &  +21 56 21 & Was 26	       & 0.063 & 1.60\pl0.07 & --0.13\pl0.04 & 0.98\pl0.02 & --0.11\pl0.06 & 
--- & --- & --- \\
50  & 11 44 29.9 &  +36 53 09 & CASG 855	       & 0.040 & 1.29\pl0.08 & --0.18\pl0.05 & --- & --- & --- & --- & --- \\
51  & 12 01 14.4 & --03 40 41 & Mkn 1310	       & 0.019 & 0.89\pl0.06 & --0.07\pl0.06 & 0.26\pl0.01 & --0.06\pl0.16 & 
--- & --- & --- \\
52  & 12 03 09.5 &  +44 31 50 & NGC 4051	       & 0.002 & 3.62\pl0.11 & --0.47\pl0.02 & 1.68\pl0.01 & --0.41\pl0.02  & 0.804 &
0.489\pl0.014 & 2.209  \\
53  & 12 04 42.1 &  +27 54 12 & GQ Com	       & 0.165 & 0.59\pl0.05 & --0.13\pl0.06 & 0.47\pl0.01 & --0.11\pl0.04 & 0.158 &
0.164\pl0.008 & 0.613  \\
54  & 12 09 45.2 &  +32 17 02 & RX J1209.8+3217  & 0.145 & 0.59\pl0.11 & --0.63\pl0.08 & 0.28\pl0.01 & --0.58\pl0.18 & 
--- & --- & --- \\
55  & 12 14 17.7 & +14 03 13 & PG1211+14         & 0.082 & 1.71\pl0.08 & --0.35\pl0.03 & 1.05\pl0.02 & --0.20\pl0.06 & 
--- & --- & --- \\
56  & 12 18 26.6 &  +29 48 46 & Mkn 766	       & 0.013 & 5.35\pl0.13 & --0.03\pl0.02 & 3.53\pl0.02 & --0.09\pl0.03  & 1.468 &
0.384\pl0.002 & 1.399  \\
57  & 12 29 06.7 &  +02 03 09 & 3C 273           & 0.158 & 7.54\pl0.16 & --0.15\pl0.02 & 6.29\pl0.03 &  +0.10\pl0.02 & 1.988 &
3.082\pl0.039 & 11.69  \\
58  & 12 31 36.6 &  +70 44 14 & RX J1231.6+7044  & 0.208 & 1.08\pl0.05 & --0.20\pl0.04 &  ---        & --- & 0.273 &
0.162\pl0.007 & 0.642  \\
59  & 12 32 03.6 &  +20 09 30 & Mkn 771	       & 0.064 & 0.61\pl0.05 & --0.39\pl0.06 & 1.32\pl0.02 & --0.26\pl0.05 & 
--- & --- & --- \\
60  & 12 33 41.7 &  +31 01 03 & CBS 150          & 0.290 & 0.51\pl0.05 & --0.53\pl0.05 & 0.33\pl0.01 & --0.53\pl0.15  &
--- & --- & --- \\
61  & 12 36 51.2 &  +45 39 05 & MCG+08-23-067     & 0.030 & 0.58\pl0.05 & --0.29\pl0.06 &  ---  & --- & --- & --- & --- \\
62  & 12 37 41.2 &  +26 42 28 & IC 3599           & 0.021 & 4.90\pl0.11 & --0.64\pl0.02 & 0.07\pl0.01 & --0.83\pl0.05   & 1.247 &
0.005\pl0.001 & 0.025  \\
63  & 12 39 39.4 & --05 20 39 & NGC 4593          & 0.009 & 3.32\pl0.18 & --0.16\pl0.05 & 1.44\pl0.04 &  +0.10\pl0.10   & 0.897 &
1.041\pl0.024 & 3.850 \\
64  & 12 42 10.6 &  +33 17 03 & IRAS F12397+3333  & 0.044 & 0.84\pl0.06 & --0.33\pl0.05 & 0.78\pl0.01 & --0.27\pl0.05   & 
--- & --- & --- \\ 
65  & 12 46 35.2 &  +02 22 09 & PG 1244+026       & 0.049 & 1.28\pl0.10 & --0.50\pl0.06 & 1.15\pl0.02 & --0.44\pl0.05  &  
--- & --- & --- \\
66  & 13 04 17.0 &  +02 05 37 & RX J1304.2+0205   & 0.229 & 0.79\pl0.07 & --0.64\pl0.05 &   ---  & --- & --- & --- & --- \\
67  & 13 09 47.0 &  +08 19 48 & PG 1307+085       & 0.155 & 0.60\pl0.06 & --0.24\pl0.06 & 0.56\pl0.01 & --0.24\pl0.06   & 0.153 &
0.100\pl0.006 & 0.395   \\
68  & 13 14 22.7 &  +34 29 39 & RX J1314.3+3429   & 0.075 & 0.68\pl0.06 & --0.62\pl0.04 & 0.74\pl0.03 & --0.70\pl0.20   & 0.128 &
0.117\pl0.005 & 0.620  \\
69  & 13 19 57.1 &  +52 35 33 & RX J1319.9+5235   & 0.092 & 0.68\pl0.05 & --0.53\pl0.05 &  ---   & --- & --- & --- & --- \\
70  & 13 23 49.5 &  +65 41 48 & PG 1322+659       & 0.168 & 0.71\pl0.05 & --0.48\pl0.04 & 0.54\pl0.01 & --0.39\pl0.07  & 
--- & --- & --- \\
71  & 13 37 18.7 &  +24 23 03 & IRAS 1334+24      & 0.108 & 2.50\pl0.09 & --0.66\pl0.03 & 1.04\pl0.02 & --0.61\pl0.09   & 0.485 &
0.319\pl0.008 & 1.644  \\
72  & 13 43 56.7 &  +25 38 48 & Ton 730           & 0.087 & 0.55\pl0.05 & --0.60\pl0.05 &  --- & --- &
--- & --- & --- \\
73  & 13 55 16.6 &  +56 12 45 & RX J1355.2+5612   & 0.122 & 0.71\pl0.06 & --0.63\pl0.04 &    ---      &  --- & 0.137 &
0.122\pl0.008 & 0.634  \\
74  & 14 05 16.2 &  +25 55 34 & PG 1402+25        & 0.164 & 0.66\pl0.05 & --0.55\pl0.05 & 0.99\pl0.02 & --0.56\pl0.10   & 0.140 &
0.337\pl0.009 & 1.590  \\
75  & 14 13 36.7 &  +70 29 50 & RX J1413.6+7029   & 0.107 & 0.76\pl0.07 & --0.16\pl0.06 &    ---       &  ---  & 0.199 &
0.105\pl0.006 & 0.401 \\
76  & 14 17 59.5 &  +25 08 12 & NGC 5548          & 0.017 & 4.76\pl0.11 & --0.14\pl0.02 & 7.95\pl0.03 & --0.19\pl0.02  & 1.270 &
0.763\pl0.019 & 2.857  \\
77  & 14 24 03.8 & --00 26 58 & QSO 1421--0013    & 0.151 & 0.58\pl0.06 & --0.06\pl0.07 &  ---   & --- & --- & --- & --- \\
78  & 14 27 25.0 &  +19 49 53 & Mkn 813           & 0.111 & 0.79\pl0.06 & --0.14\pl0.05 &   ---  & --- & --- & --- & --- \\
79  & 14 31 04.1 &  +28 17 14 & Mkn 684           & 0.046 & 0.53\pl0.06 & --0.24\pl0.07 &    ---      &   ---  & 0.131 &
0.247\pl0.010 & 1.002 \\
80  & 14 42 07.5 &  +35 26 23 & Mkn 478           & 0.077 & 5.31\pl0.11 & --0.73\pl0.01 & 3.17\pl0.05 & --0.75\pl0.09  & 0.958 &
0.119\pl0.004 & 0.660  \\
\noalign{\smallskip}\hline
\end{tabular}
\end{table}
}

\newpage
\clearpage

{\small
\begin{table}
\hspace*{-1.5cm}
\begin{tabular}{rlllcccccccc}
\hline\noalign{\smallskip}
& & & & & \multicolumn{2}{c}{RASS} & \multicolumn{2}{c}{PSPC pointed} & \multicolumn{3}{c}{HRI pointed} \\
\rb{No.}  & \rb{$\alpha_{2000}$} & \rb{$\delta_{2000}$} & \rb{Name}  & \rb{z} & 
CR & HR1 & CR & HR1  & expected & CR & converted  \\
& h m s & $^{\circ}~^{'}~^{''}$ & & & $\rm cts~s^{-1}$ & & $\rm cts~s^{-1} $ &  & HRI CR & \cts & CR \\
\noalign{\smallskip}\hline\noalign{\smallskip}
81  & 14 51 08.8 &  +27 09 27 & PG 1448+273       & 0.065 & 0.78\pl0.06 & --0.03\pl0.05 &   ---  & --- & --- & --- & --- \\
82  & 15 04 01.2 &  +10 26 16 & Mkn 841           & 0.036 & 0.77\pl0.08 & --0.18\pl0.06 & 2.20\pl0.01 & --0.07\pl0.02  & 0.205 &
0.888\pl0.019 & 3.303 \\
83  & 15 29 07.5 &  +56 16 07 & HS 1529+56        & 0.100 & 0.79\pl0.05 & --0.46\pl0.03 &   ---  & --- & --- & --- & --- \\
84  & 15 59 09.7 &  +35 01 48 & Mkn 493           & 0.032 & 0.55\pl0.06 & --0.24\pl0.07 & 0.40\pl0.01 & --0.23\pl0.07 
--- & --- & --- \\
85  & 16 13 57.2 &  +65 43 11 & Mkn 876           & 0.129 & 0.84\pl0.05 & --0.08\pl0.03 & 0.98\pl0.01 & --0.04\pl0.05  &
--- & --- & --- \\
86  & 16 18 09.4 &  +36 19 58 & RX J1618.1+3619   & 0.034 & 0.87\pl0.05 & --0.44\pl0.03 &    ---      &   ---  & 0.193 &
0.113\pl0.005 & 0.509 \\
87  & 16 19 51.3 &  +40 58 48 & KUG 1618+40       & 0.038 & 0.57\pl0.06 & --0.56\pl0.04 &   ---  & --- &  --- & --- & --- \\
88  & 16 24 56.6 &  +75 54 56 & RX J1624.9+7554   & 0.065 & 0.44\pl0.04 & --0.22\pl0.04 &    $^3$      &    $^3$      & 
--- & --- & --- \\
89  & 16 27 56.1 &  +55 22 32 & PG 1626+554       & 0.133 & 0.61\pl0.05 & --0.37\pl0.04 & 1.05\pl0.02 & --0.49\pl0.09  & 
--- & --- & --- \\
90  & 16 29 01.3 &  +40 08 00 & EXO 1627+4014     & 0.272 & 0.84\pl0.05 & --0.80\pl0.02 & 0.99\pl0.01 & --0.77\pl0.09  &  
--- & --- & --- \\
91  & 17 02 31.1 &  +32 47 20 & HS 1702+32        & 0.164 & 0.58\pl0.05 & --0.48\pl0.04 & 0.81\pl0.01 & --0.62\pl0.07  & 
--- & --- & --- \\
92  & 21 32 27.9 &  +10 08 20 & II Zw 136         & 0.065 & 1.30\pl0.07 &  +0.01\pl0.04 & 1.16\pl0.02 &  +0.08\pl0.04  & 0.393 &
0.265\pl0.011 & 0.874 \\
93  & 21 46 36.0 & --30 51 41 & RX J2146.6--3051  & 0.075 & 0.77\pl0.06 & --0.14\pl0.06 &   ---  & --- & --- & --- & --- \\
94  & 22 07 45.0 & --32 35 01 & A 09.25           & 0.063 & 0.98\pl0.07 & --0.25\pl0.06 &   ---  & --- &  --- & --- & --- \\
95  & 22 09 07.6 & --27 48 36 & NGC 7214          & 0.023 & 0.78\pl0.08 & --0.25\pl0.07 & 2.69\pl0.01 & --0.40\pl0.02  & 0.198 &
0.133\pl0.02 & 0.525  \\
96  & 22 16 53.2 & --44 51 57 & RX J2216.8--4451  & 0.136 & 1.39\pl0.08 & --0.66\pl0.03  &   ---  & --- & 0.279 & 
0.140\pl0.005 & 0.698 \\
97  & 22 17 56.6 & --59 41 30 & RX J2217.8--5941  & 0.160 & 0.83\pl0.06 & --0.67\pl0.04 &    ---   &   ---  & 0.169 & 
0.005\pl0.002 & 0.026 \\
98  & 22 30 40.3 & --39 42 52 & PKS 2227-399      & 0.318 & 0.55\pl0.06 & --0.10\pl0.08 & 0.61\pl0.01 & --0.19\pl0.08  & 
--- & --- & --- \\
99  & 22 42 37.7 & --38 45 16 & RX J2242.6--3845  & 0.221 & 0.65\pl0.06 & --0.70\pl0.05 &    ---    &   ---  & 0.117 &
0.060\pl0.010 & 0.333  \\
100 & 22 45 20.3 & --46 52 12 & RX J2245.2--4652  & 0.201 & 1.60\pl0.09 & --0.70\pl0.04 & 1.98\pl0.02 & --0.71\pl0.05  & 0.305 &
0.218\pl0.007 & 1.147  \\
101 & 22 48 41.2 & --51 09 53 & RX J2248.6--5109  & 0.102 & 2.58\pl0.10 & --0.61\pl0.03 & 2.41\pl0.02 & --0.53\pl0.05  & 
--- & --- & --- \\
102 & 22 57 39.0 & --36 56 07 & MS 2254-36        & 0.039 & 2.04\pl0.11 & --0.56\pl0.04 & 2.29\pl0.03 & --0.61\pl0.06  & 
--- & --- & --- \\
103 & 22 58 45.4 & --26 09 14 & RX J2258.7--2609  & 0.076 & 0.66\pl0.09 & --0.16\pl0.12 &    ---  &   ---  & 0.128 & 
0.122\pl0.008 & 0.625  \\
104 & 23 01 36.2 & --59 13 20 & RX J2301.6--5913  & 0.149 & 0.91\pl0.07 & --0.12\pl0.06 &   ---  & --- & --- & --- & --- \\
105 & 23 01 52.0 & --55 08 31 & RX J2301.8--5508  & 0.140 & 0.90\pl0.07 & --0.63\pl0.04 & 0.65\pl0.01 & --0.67\pl0.07  & 
--- & --- & --- \\
106 & 23 04 37.3 & --35 01 13 & RX J2304.6--3501  & 0.042 & 0.61\pl0.08 & --0.47\pl0.09 & 0.19\pl0.01 & --0.48\pl0.10  & 
--- & --- & --- \\
107 & 23 12 34.8 & --34 04 20 & RX J2312.5--3404  & 0.202 & 0.43\pl0.08 & --0.20\pl0.12 &  ---   & --- & --- & --- & --- \\
108 & 23 17 49.9 & --44 22 28 & RX J2317.8--4422  & 0.132 & 0.80\pl0.08 & --0.79\pl0.05 &    ---  &   ---  & 0.142 & 
0.072\pl0.005 & 0.410  \\
109 & 23 25 11.8 & --32 36 35 & RX J2325.2--3236  & 0.216 & 0.58\pl0.10 & --0.62\pl0.11 &    ---  &   ---  & 0.115 & 
0.085\pl0.005 & 0.425  \\
110 & 23 25 24.2 & --38 26 49 & IRAS 23226-38     & 0.036 & 1.27\pl0.12 & --0.11\pl0.09 &   ---  & --- &  --- & --- & --- \\
111 & 23 43 28.6 & --14 55 30 & MS 23409-1511     & 0.137 & 0.81\pl0.07 & --0.44\pl0.07 & 0.89\pl0.01 & --0.46\pl0.04  &  
--- & --- & --- \\
112 & 23 49 24.1 & --31 26 03 & RX J2349--3126    & 0.135 & 0.64\pl0.06 & --0.53\pl0.06 & 0.11\pl0.01 & --0.21\pl0.13  & 
--- & --- & --- \\
113 & 23 57 28.0 & --30 27 40 & AM 2354-304       & 0.033 & 0.56\pl0.06 & --0.26\pl0.07 &   ---  & --- & --- & --- & --- \\
\noalign{\smallskip}\hline \\
\end{tabular}

$^1$ WPVS007: Because of the small number of photons no hardness ratio is given. \\
$^2$ IC 3599: Merged event file of ROR 702704 and 702706 \\
$^3$ RX J1624+75: Only upper limits are available (Grupe et al. 1999b)

\end{table}
}

\newpage
\clearpage

{\small
\begin{table}
\caption{\label{obs_list} List of observations, RASS and pointed PSPC and HRI.
Listed are begin and end of the observing
period, the total exposure time, and the ROSAT Observation Request number (ROR) 
}
\begin{tabular}{rlccrccrlccrl}
\hline\noalign{\smallskip}
 & & \multicolumn{3}{c}{RASS} & \multicolumn{4}{c}{PSPC-pointed} & \multicolumn{4}{c}{HRI pointed} \\
\rb{No.} & \rb{name}  & begin & end & $\rm T_{obs}$ & begin & end & $\rm T_{obs}$ &  & 
begin & end & $\rm T_{obs}$ &  \\
& & yymmdd.d & yymmdd.d   & s & yymmdd.d & yymmdd.d & s & \rb{ROR} & yymmdd.d & yymmdd.d & s & \rb{ROR} \\
\noalign{\smallskip}\hline\noalign{\smallskip}
        1 & Mkn 335      & 900713.8 & 900715.2 &  375  & 910629.8 & 910630.8 & 24337 & 700101 &
		971220.4 & 971221.3 & 11014 & 601123 \\
        2 & ESO 242-G8   & 901125.4 & 901127.3 &  238  & 930604.4 & 930605.8 & 6208  & 701148 &
	        951225.5 & 951225.5 & 2291 & 702639 \\
        3 & WPVS 007     & 901123.2 & 901125.4 &  299  & 931111.5 & 931113.5 & 9691  & 701527 & 
	        960613.4 & 960624.7 & 19863 & 702705 \\
        4 & RX J0057-22  & 901216.9 & 901218.5 &  466  & 920618.6 & 920618.7 & 2961  & 701141 &
	        --- & --- & --- & --- \\ 
        5 & QSO 0056-36  & 901209.4 & 901211.5 &  528  & --- & --- & --- & --- & 
	        961221.0 & 961230.0 & 2256  & 702905 \\ 
        6 & RX J0100-51  & 901128.0 & 901130.2 &  306  & --- & --- & --- & --- & 
	        960602.1 & 960625.4 & 5155 & 702633 \\ 
        7 & RX J0105-14  & 900712.9 & 900714.8 &  579  & --- & --- & --- & --- & --- & --- & --- & --- \\
        8 & RX J0117-38  & 901212.0 & 901214.4 &  540  & --- & --- & --- & --- & --- & --- & --- & --- \\
        9 & MS 0117-28   & 901218.8 & 901220.9 &  412  & 911228.5 & 911229.3 & 4467  & 700445 & --- & --- & --- & --- \\ 
       10 & RX J0128-18  & 900715.9 & 910117.0 &  418  & --- & --- & --- & --- & --- & --- & --- & --- \\ 
       11 & IRAS01267-21 & 900715.2 & 910116.2 &  344  & --- & --- & --- & --- & 960112.4 & 960112.5 & 4853 & 702655 \\
       12 & RX J0148-27  & 900715.9 & 910117.2 &  504  & 920609.4 & 920610.1 & 6682 & 701185 & --- & --- & --- & ---  \\
       13 & RX J0152-23  & 901229.5 & 901231.2 &  608  & --- & --- & ---  & --- & 960101.8 & 960112.3 & 4978 & 702654 \\
       14 & Mkn 1044     & 910114.3 & 910119.3 &  370  & 920809.5 & 920810.5 & 2836 & 700792 & --- & --- & --- & ---  \\
       15 & Mkn 1048     & 910119.0 & 910120.4 &  382  & 930809.2 & 930810.0 & 6191 & 701599 & 
           960115.8 & 960115.0 & 2411 & 600866 \\
       16 & RX J0311-20  & 910122.9 & 910124.3 &  319  & --- & --- & --- & --- & ---  & --- & --- & --- \\
       17 & RX J0319-26  & 910122.9 & 910124.3 &  263  & 920818.8 & 920821.3 & 6823 & 701052 & --- & --- & --- & ---  \\
       18 & RX J0323-49  & 910102.4 & 910105.7 &  745  & 921222.9 & 921223.9 & 5914 & 701183 & 
           960817.7 & 960819.7 & 2428 & 702670 \\
       19 & ESO 301-G13  & 910111.4 & 910114.7 &  657  & --- & --- & ---  & --- & 960102.7 & 960105.0 & 3805 & 702657 \\
       20 & VCV0331-37   & 910121.0 & 910123.1 &  244  & --- & --- & ---  & --- & 970713.3 & 970713.4 & 3427 & 702902 \\
       21 & RX J0349-47  & 910113.7 & 910120.2 &  421  & --- & --- & ---  & --- & 960117.4 & 960117.5 & 2049 & 600869 \\
       22 & Fair 1116    & 910124.0 & 910125.2 &  167  & --- & --- & ---  & --- & 970712.6 & 970712.6 & 2754 & 702903 \\
       23 & CTS 44       & 910805.4 & 910810.8 &  588  & --- & --- & ---  & --- & --- & --- & --- & --- \\
       24 & RX J0412-47  & 910124.0 & 910125.2 &  163  & --- & --- & ---  & --- & 951230.0 & 951230.2 & 5367 & 702662 \\
       25 & RX J0426-57  & 910109.1 & 910119.3 &  796  & 920407.3 & 920407.8 & 4094  & 700034 &
           960816.5 & 960816.6 & 4703 & 702038 \\
       26 & Fair 303     & 910122.1 & 910125.2 &  290  & 911220.4 & 911220.5 & 2669 & 800043 & --- & --- & --- & ---  \\
       27 & RX J0437-47  & 900730.6 & 910812.4 & 1037  & 920920.7 & 920921.8 & 6142 & 701184 &
           970922.1 & 970926.1 & 23114 & 400897  \\
       28 & RX J0439-53  & 910805.3 & 910809.8 &  148  & 970220.6 & 970220.7 & 1370 & 190537 & --- & --- & --- & ---  \\
       29 & H 0439-272   & 900812.6 & 900814.6 &  122  & 910310.2 & 910310.3 & 4955 & 700035 & --- & --- & --- & ---  \\
       30 & 1 ES 0614-58 & 900920.7 & 901003.2 & 2792  & --- & --- & ---  & --- & --- & --- & --- & --- \\
       31 & RX J0859+48  & 901014.9 & 901017.0 &  448  & --- & --- & ---  & --- & --- & --- & --- & --- \\
       32 & RX J0902-07  & 901102.5 & 901104.2 &  410  & --- & --- & ---  & --- & --- & --- & --- & --- \\
       33 & Mkn 110      & 901018.2 & 901020.0 &  617  & 911101.8 & 911102.8 & 6013 & 700262 & 
          951016.8 & 951101.9 & 30308 & 702627 \\
       34 & PG 0953+41   & 901029.4 & 901031.3 &  485  & 920430.2 & 920430.4 & 7014 & 700526 & --- & --- & --- & ---  \\
       35 & RX J1005+43  & 901030.1 & 901101.0 &  686  & --- & --- & --- & --- & 951101.5 & 951108.4 & 4711 & 702656  \\
       36 & RX J1007+22  & 901108.4 & 901110.1 &  599  & --- & --- & --- & --- & --- & ---  & --- & --- \\
       37 & CBS 126      & 901104.1 & 901105.9 &  642  & --- & --- & --- & --- & 951030.8 & 951031.1 & 3901 & 702651 \\
       38 & HS 1019+37   & 901104.4 & 901106.3 &  643  & 931029.7 & 931030.2 & 13510 & 800491 & --- & --- & --- & --- \\
       39 & Mkn 141      & 901019.2 & 901021.9 &  694  & --- & --- & --- & --- & 961010.4 & 961010.8 & 3589 & 702664 \\
       40 & Mkn 142      & 901029.3 & 901031.5 &  762  & 931021.2 & 931022.6 & 8427 & 600624 & --- & --- & ---  & --- \\
\noalign{\smallskip}\hline
\end{tabular}
\end{table}
}

\newpage
\clearpage

{\small
\begin{table}
\begin{tabular}{rlccrccrlccrl}
\hline\noalign{\smallskip}
 & & \multicolumn{3}{c}{RASS} & \multicolumn{4}{c}{PSPC-pointed} & \multicolumn{4}{c}{HRI pointed} \\
\rb{No.} & \rb{name}  & begin & end & $\rm T_{obs}$ & begin & end & $\rm T_{obs}$ &  & 
begin & end & $\rm T_{obs}$ &  \\
& & yymmdd.d & yymmdd.d   & s & yymmdd.d & yymmdd.d & s & \rb{ROR} & yymmdd.d & yymmdd.d & s & \rb{ROR} \\
\noalign{\smallskip}\hline\noalign{\smallskip}
       41 & RX J1034+39  & 901106.8 & 901108.8 &  525   & 911118.8 & 911119.1 & 4617 & 700551 & --- & --- & --- & ---  \\
       42 & RX J1117+65  & 901025.8 & 901028.8 &  874   & --- & --- & --- & --- & 961022.4 & 961022.4 & 2169 & 702892 \\
       43 & PG 1115+407  & 901115.2 & 901117.3 &  446   & 930520.4 & 930520.8 & 6489 & 700801 &
           951114.9 & 951115.0 & 3645  & 702191 \\
       44 & Ton 1388     & 901124.8 & 901126.5 &  422   & 910529.3 & 910530.2 & 24294 & 700228 & 
           950528.0 & 950528.1 & 1774 & 701690 \\
       45 & EXO 1128+69  & 901023.1 & 901026.3 & 1099   & --- & --- & --- & --- & --- & --- & --- & --- \\
       46 & B2 1128+31   & 901122.8 & 901124.4 &  423   & --- & --- & --- & --- & --- & --- & --- & --- \\
       47 & SBS 1136+579 & 901106.2 & 901108.9 &  491   & 930530.1 & 930530.3 & 5018 & 800402 & --- &  --- & --- & ---  \\
       48 & Z 1136+3412  & 901123.2 & 901125.1 &  422   & 930518.5 & 930519.1 & 2046 & 201120 & --- & --- & --- & ---  \\
       49 & Was 26       & 901129.5 & 901201.3 &  602   & 921207.0 & 921207.5 & 4889 & 701149 & --- & --- & --- & ---  \\
       50 & CASG 855     & 901122.8 & 901124.3 &  378   & --- & --- & --- & --- & --- & --- & --- & --- \\
       51 & Mkn 1310     & 901214.8 & 901216.5 &  481   & 920707.8 & 920707.9 & 3341 & 201367 & --- & --- & --- & --- \\
       52 & NGC 4051     & 901121.8 & 901123.9 &  506   & 911116.0 & 911116.9 & 28459 & 700557 & 
           950520.5 & 950520.6 & 2587 & 701884 \\
       53 & GQ Com       & 901201.9 & 901203.8 &  666   & 910604.2 & 910605.2 & 25991 & 700232 & 
           960621.4 & 960621.5 & 4639 & 702171 \\
       54 & RX J1209+32  & 901130.7 & 901202.7 &  210   & 930523.5 & 930523.8 & 2748 & 701051 & --- & --- & --- & --- \\
       55 & PG 1211+143  & 901210.5 & 901212.3 &  609   & 911217.1 & 911224.7 & 3803 & 700018 & --- & --- & --- & --- \\
       56 & Mkn 766      & 901203.9 & 901205.9 &  525   & 921208.6 & 921210.2 & 6774 & 701091 & 
           960621.7 & 960621.7 & 1518  & 702633 \\
       57 & 3C 273       & 901218.9 & 901220.7 &  496   & 911214.3 & 911215.2 & 6140 & 700191 & 
           941222.1 & 941222.1 & 2088  & 701961 \\
       58 & RX J1231+70  & 901026.4 & 901030.2 & 1005   & --- & --- & --- & --- & 960504.9 & 960505.7 & 4285 & 702652 \\
       59 & Mkn 771      & 901211.9 & 901213.7 &  604   & 911213.3 & 911228.2 & 6402 & 700435 & --- & --- & --- & --- \\
       60 & CBS 150      & 901206.7 & 901208.8 &  690   & 920702.3 & 920702.7 & 3120 & 701050 & --- & --- & --- & --- \\
       61 & MCG+08-23-06 & 901127.7 & 901130.1 &  661   & --- & --- & --- & --- & --- & --- & --- & --- \\
       62 & IC 3599      & 901209.9 & 901211.9 &  658   & 920630.3 & 920701.9 & 5422 &701097 & 
          960630.5 & 960702.1 & 31621  & $^1$ \\
       63 & NGC 4593     & 900715.1 & 910115.8 &  180   & 920714.3 & 920714.3 & 1261 & 701012 & 
          960114.9 & 960116.4 & 1912 & 701885 \\
       64 & IRAS 1239+33 & 901207.3 & 901209.3 &  707   & 911215.7 & 920104.2 & 19536 & 600129 & --- & --- & --- & --- \\
       65 & PG 1244+026  & 900714.1 & 900715.3 &  239   & 911222.2 & 911224.8 & 5486 & 700020 & --- & --- & --- & --- \\
       66 & RX J1304+02  & 901227.9 & 910117.2 &  401   & --- & --- & --- & --- & --- & --- & --- & --- \\
       67 & PG 1307+085  & 900716.2 & 910117.0 &  509   & 920713.5 & 920719.0 & 7682 & 700229 & 
           950714.7 & 950714.9 & 2950 & 400625 \\
       68 & RX J1314+34  & 901213.9 & 901216.1 &  740   & 920706.8 & 920706.8 & 2526 & 700802 & 
           960108.8 & 960110.1 & 5916  & 702660 \\
       69 & RX J1319+52  & 901129.8 & 901202.8 &  714   & --- & --- & --- & --- & --- & --- & --- & --- \\
       70 & PG 1322+659  & 901110.9 & 901114.5 &  810   & 921130.0 & 921201.9 & 8387 & 700803 & --- & --- & --- & --- \\
       71 & IRAS 1334+24 & 910115.3 & 910116.8 &  506   & 920107.7 & 920114.6 & 3281 & 700553 & 
           960716.7 & 960721.8 & 6207 & 702625 \\
       72 & Ton 730      & 900716.3 & 910117.1 &  592   & --- & --- & --- & --- & --- & --- & --- & --- \\
       73 & RX J1355+56  & 901202.1 & 901205.6 &  829   & --- & --- & --- & --- & 970529.8 & 970529.9 & 2577 & 702897 \\
       74 & PG 1402+25   & 901231.4 & 910102.6 &  627   & 920107.7 & 920113.3 & 2692 & 700226  &
           960114.2 & 960114.3 & 4379  & 702152 \\
       75 & RX J1413+70  & 901104.2 & 901109.2 &  501   & --- & --- & --- & --- & 951023.9 & 951114.3 & 4140 & 702659 \\
       76 & NGC 5548     & 910104.1 & 910117.5 &  631   & 921225.4 & 921225.7 & 8126 & 700536 & 
           950124.6 & 950125.1 & 2260  & 701950 \\
       77 & Q 1421-0013  & 910119.5 & 910120.8 &  442   & --- & --- & --- & --- & --- & --- & --- & --- \\
       78 & Mkn 813      & 910109.2 & 910111.1 &  684   & --- & --- & --- & --- & --- & --- & --- & --- \\
       79 & Mkn 684      & 910105.8 & 910118.7 &  538   & --- & --- & --- & --- & 970719.9 & 970720.1 & 2835 & 702907 \\
       80 & Mkn 478      & 910104.2 & 910118.0 &  692   & 920117.6 & 920117.8 & 2354 & 700559	& 
           971224.1 & 971224.5 & 9632 & 704273  \\
\noalign{\smallskip}\hline
\end{tabular}
\end{table}
}

\newpage
\clearpage

{\small
\begin{table}
\begin{tabular}{rlccrccrlccrl}
\hline\noalign{\smallskip}
 & & \multicolumn{3}{c}{RASS} & \multicolumn{4}{c}{PSPC-pointed} & \multicolumn{4}{c}{HRI pointed} \\
\rb{No.} & \rb{name}  & begin & end & $\rm T_{obs}$ & begin & end & $\rm T_{obs}$ &  & 
begin & end & $\rm T_{obs}$ &  \\
& & yymmdd.d & yymmdd.d   & s & yymmdd.d & yymmdd.d & s & \rb{ROR} & yymmdd.d & yymmdd.d & s & \rb{ROR} \\
\noalign{\smallskip}\hline\noalign{\smallskip}
       81 & PG 1448+273  & 910111.8 & 910114.1 &  701   & --- & --- & --- & --- & --- & --- & --- & ---\\
       82 & Mkn 841      & 910124.4 & 910125.8 &  451   & 920120.2 & 920126.1 & 16842 & 700257 &
           950801.0 & 950808.9 & 2699 & 701677 \\
       83 & HS 1529+56   & 900714.2 & 910117.1 & 1287   & --- & --- & --- & --- & --- & --- & --- & ---\\
       84 & Mkn 493      & 910807.3 & 910810.0 &  553   & 920201.4 & 920129.6 & 8038 & 700096 & --- & --- & --- & ---  \\
       85 & Mkn 876      & 901201.6 & 901210.8 & 1906   & 911123.3 & 911203.9 & 6514 & 700230 & --- & --- & --- & --- \\
       86 & RX J1618+36  & 900730.6 & 910803.1 & 1142   & --- & --- & --- & --- & 960220.9 & 960222.8 & 4473 & 702658 \\
       87 & KUG 1618+40  & 900730.3 & 910812.5 & 1046   & --- & --- & --- & --- & --- & --- & --- & --- \\
       88 & RX J1624+75  & 901007.4 & 901015.9 & 1766   & 920112.9 & 920113.8 & 2923 & 141829 & --- & --- & --- & --- \\
       89 & PG 1626+554  & 910114.5 & 910123.2 & 1122   & 930803.5 & 930804.7 & 2328 & 701372 & --- & --- & --- & --- \\
       90 & EXO 1627+401 & 900801.2 & 900804.9 & 1188   & 930730.2 & 930730.8 & 5175 & 701507 & --- & --- & --- & --- \\
       91 & HS 1702+32   & 900818.9 & 910218.1 & 1117   & 930803.3 & 930803.7 & 11317 & 800530 & --- & --- & --- & --- \\
       92 & II Zw 136    & 901111.5 & 901113.3 &  562   & 930529.8 & 930605.1 & 8622 & 701252 &
          951117.2 & 951117.2 & 2155 & 702638  \\
       93 & RX J2146-30  & 901031.5 & 901102.3 &  475   & --- & --- & --- & --- & --- & --- & --- & --- \\
       94 & A 09.25      & 901104.4 & 901106.1 &  342   & --- & --- & --- & --- & --- & --- & --- & --- \\
       95 & NGC 7214     & 901106.6 & 901108.4 &  328   & 920501.1 & 920502.2 & 15756 & 600175 &
           951113.0 & 951114.8 & 34040 & 800848 \\
       96 & RX J2216-44  & 901031.6 & 901102.6 &  471   & --- & --- & --- & --- & 951108.4 & 951108.4 & 57 & 702650 \\
       97 & RX J2217-59  & 901022.5 & 901024.8 &  587   & --- & --- & --- & --- & 980415.7 & 980418.0 & 9760 & 601124 \\
       98 & PKS 2227-399 & 901106.1 & 901107.6 &  425   & 930530.7 & 930530.8 & 3652 & 701081 & --- & --- & --- & --- \\
       99 & RX J2242-38  & 901109.1 & 901110.7 &  472   & --- & --- & --- & --- & 961201.8 & 961204.4 & 756 & 702895 \\
      100 & RX J2245-46  & 901105.2 & 901107.1 &  340   & 921118.9 & 921128.6 & 5507 & 701186 & 
           941129.4 & 941130.8 & 4515  & 201870 \\
      101 & RX J2248-51  & 901103.1 & 901104.7 &  499   & 931102.3 & 931103.0 & 4517 & 701601 & --- & --- & --- & --- \\
      102 & MS 2254-36   & 901113.0 & 901114.3 &  284   & 920502.8 & 920502.9 & 5514 & 700554 & --- & --- & --- & --- \\
      103 & RX J2258-26  & 901118.1 & 901120.0 &  143   & --- & --- & --- & --- & 961117.5 & 961118.4 & 2538 & 702909 \\
      104 & RX J2301-59  & 901029.8 & 901101.3 &  485   & --- & --- & --- & --- & --- & --- & --- & --- \\
      105 & RX J2301-55  & 901102.3 & 901104.7 &  541   & 930518.7 & 930531.5 & 8806 & 701189 & --- & --- & --- & --- \\
      106 & RX J2304-35  & 901115.3 & 901117.2 &  204   & 921201.7 & 921204.3 & 9101 & 701187 & --- & --- & --- & --- \\
      107 & RX J2312-34  & 901117.9 & 901119.0 &  175   & --- & --- & --- & --- & --- & --- & --- & --- \\
      108 & RX J2317-44  & 901113.0 & 901114.3 &  269   & --- & --- & --- & --- & 970507.3 & 970507.4 & 3050 & 702889$^2$  \\
      109 & RX J2325-32  & 901121.1 & 901122.1 &  104   & --- & --- & --- & --- & 961130.7 & 961130.8 & 3953 & 702896 \\
      110 & IRAS 23226-3 & 901118.0 & 901119.2 &  142   & --- & --- & --- & --- & --- & --- & --- & --- \\
      111 & MS 23409-151 & 901203.1 & 901204.7 &  341   & 921220.7 & 921221.6 & 14428 & 701205 & --- &  --- & --- & ---\\
      112 & RX J2349-31  & 901126.7 & 901128.7 &  452   & 930531.1 & 930531.3 & 8148 & 701190 & --- & --- & --- & --- \\
      113 & AM 2354-304  & 901129.1 & 901130.6 &  446   & --- & --- & --- & --- & --- & --- & --- & --- \\
\noalign{\smallskip}\hline
\end{tabular}

$^1$ IC 3599: Merged event file of ROR 702704 and 702706 \\
$^2$ RX J2317--44: ROR = 702889h-1, 702889 is empty

\end{table}
}

\end{landscape}

 Fig. \ref{cr-cr} shows 
that most sources show variability on long timescales of years by factors of
about 2-3 (see also Table \ref{CR_list}), 
but do not show dramatic changes in their count rates. However four sources 
are highly 
variable in X-rays. Three of these sources are
considered to be X-ray transient AGN: RX J1624.9+7554 (Grupe et al. 1999b),
 WPVS007 (Grupe et al. 1995b), 
 and  IC 3599 (Brandt et al. 1995, Grupe et al. 1995a).  The other source, RX
 J2217.9--5941, has changed its count rate between RASS and pointed observations by
 a factor of more than 30  (Grupe et al. 2000b, and see below). 
  We can exclude
 line-of-sight Cataclysmic Variables in our Galaxy as the origin of the
 transience (see discussions in the separate papers of the sources). 
Four
sources have become brighter in their pointed PSPC observation by a factor of 
more
than 2. The sources with the highest count rate ratios are Mkn 771, Mkn 841, 
NGC 7214, and  Mkn 110.

\begin{figure*}[t]
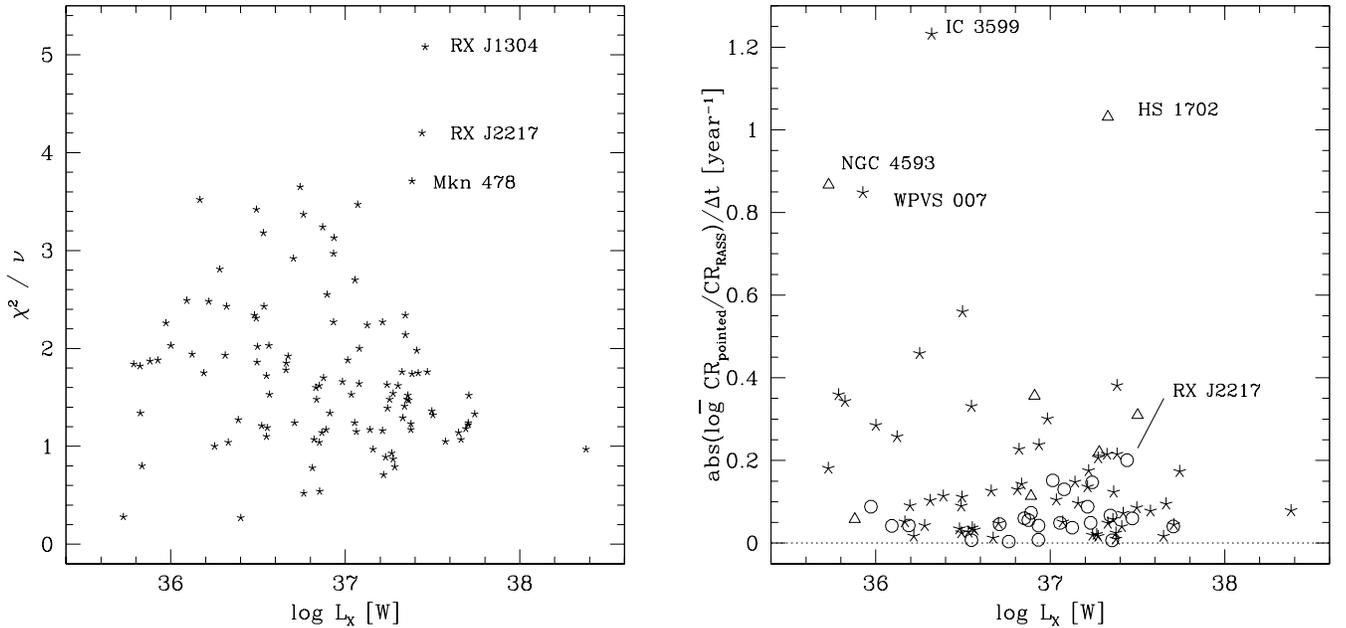

\chartlinea{H2394_F2}{H2394_F3}
\caption[ ]{\label{var_check} Variability checks for the soft X-ray AGN sample;
the left panel shows the short-term variability in the RASS data
and the right panel the long-term variability for all objects for which more 
than one observation exists (see text).
(RASS vs. PSPC observations are marked as
stars, RASS vs. HRI as open circles, and RASS vs RASS as open triangles). 
}
\end{figure*}

We performed two tests for  variability, a $\chi^2$ test 
for the RASS data (short term
behavior) and a check on the amplitude change per unit time for the long term
behavior. Fig \ref{var_check} displays the results of 
variability tests as a function of the
X-ray luminosity in the rest-frame energy 0.2-2.0 keV.  
In the left panel of Fig. \ref{var_check} the short term variability of the RASS
data is displayed (NGC 4051 and Mkn 766 are off the plot).
For objects which have RASS observations that were about half a year apart
(see Table \ref{rass_rass}), 
only the one with the longest exposure was used. A RASS coverage was typically a
few days (see Table \ref{obs_list} for details) with an time increment of 96
minutes (the ROSAT orbit). 
The reduced $\chi^2$ 
was calculated from the  mean count rate taking the count rate errors into
account.
We see that there is a slight dependence of the variability strength on
luminosity. Low-luminosity sources tend to be slightly more variable than
high-luminosity sources. However, Mkn 478 and RX J1304+02, and RX J2217--59
fall off this trend. The two sources with the strongest variability on short 
timescales are NGC 4051 and Mkn 766, both well-known for their strong 
variability (see e.g. Peterson et al. 2000 and Leighly et al. 1996). 
 The other two sources with relatively strong variability, RX J1304+02 and
RX J2217--59  are `new'. On long timescales, RX J1304+02 lacks in
further pointed observations and RX J2217--59 is a
transient candidate (Grupe et al. 2000b and see below).
The X-ray transient AGN do not show unusually
strong variability on short timescales.

\begin{table*}
\begin{flushleft}
\caption{\label{rass_rass} List of sources that were observed twice during
the RASS half a year apart. Observing dates give the beginning of the RASS
coverage in yymmdd, the exposure time T$_{\rm exp}$ is given in s, and the
count rate in \cts.
 }
\begin{tabular}{rlrrcrrc}
\hline\noalign{\smallskip}
& & \multicolumn{3}{c}{first observation} & \multicolumn{3}{c}{second
observation} \\
\rb{\#} & \rb{name} & UT date & T$_{\rm exp}$ & CR & UT date & T$_{\rm exp}$ & CR \\
\noalign{\smallskip}\hline
10 & RX J0128-18  & 900715 &  99.4 & 0.088\pl0.130 & 910115 & 318.5 & 1.056\pl0.434 \\  
11 & IRAS01267-21 & 900715 & 210.0 & 0.724\pl0.259 & 910115 & 133.8 & 0.837\pl0.409 \\ 
12 & RX J0148-27  & 900715 &  89.4 & 1.760\pl0.356 & 901227 & 414.1 & 2.515\pl0.538 \\
63 & NGC 4593     & 900715 &  50.1 & 1.440\pl0.663 & 910116 & 130.2 & 3.908\pl0.373 \\ 
67 & PG 1307+085  & 900716 &  66.0 & 0.506\pl0.264 & 910115 & 443.1 & 0.652\pl0.261 \\ 
83 & HS 1529+56   & 900714 & 583.6 & 0.963\pl0.293 & 901227 & 703.7 & 0.639\pl0.261 \\ 
87 & KUG 1618+40  & 900730 & 473.9 & 0.527\pl0.247 & 910810 & 572.2 & 0.602\pl0.331 \\ 
91 & HS 1702+32   & 900818 & 810.6 & 0.345\pl0.201 & 910217 & 306.7 & 1.130\pl0.406 \\ 
\noalign{\smallskip}\hline
\end{tabular}
\end{flushleft}

\end{table*}

The long-term behaviour is shown in the right panel of Fig. \ref{var_check}. We
also see a slight
tendency for low-luminosity sources to be more variable.
We took all sources into account for which at least two
observations existed that were more than half a year apart. This includes also 
those sources for which two RASS coverages existed (see Table 
\ref{rass_rass}, marked as triangles
in Fig. \ref{var_check}). 
Clearly, in the long-term behaviour, the transients have
the highest change in amplitude over time. Two other sources, NGC 4593 and 
HS1702+32
also show very strong variability on long-timescales. Note that our transient 
candidate RX J2217.9--5941 is the source with the strongest variability between 
RASS and HRI observation.

\begin{figure}
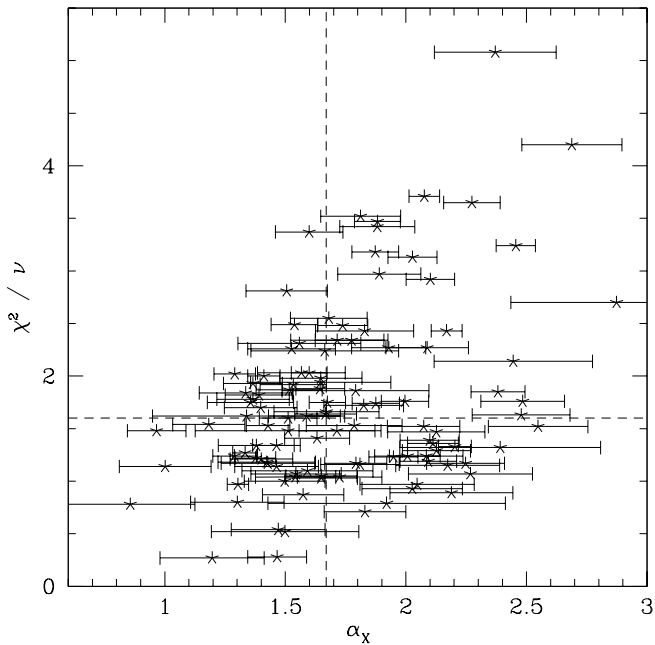

\clipfig{H2394_F4}{87}{08}{53}{200}{248}
\caption[ ]{\label{ax_chisq} Short-term
$\chi^2$ variability  vs. X-ray spectral 
slope \ax.
The dashed lines mark the medians in \ax~ and $\chi^2/\nu$. Three objects are 
off the plot, NGC 4051, Mkn 766 and WPVS007 (with positions at 1.62,18.54; 
1.11,8.86; and 8.70,1.88 respectively).
}
\end{figure}

Fig. \ref{ax_chisq} displays the variability strength $\chi^2/\nu$ 
(same as in Fig. \ref{var_check}) as a function of the
X-ray spectral slope \ax. We found that sources with steep X-ray spectra 
tend to show stronger variabilities. The dashed lines in Fig. \ref{ax_chisq} 
mark the medians of \ax~ and $\chi^2/\nu$ without the high variability 
sources NGC 4051 and Mkn 766 and the source with the steepest X-ray spectrum,
WPVS007. Dividing our sample in a steep and flat X-ray spectrum (divided by the 
median \ax=1.67) we find a mean $\chi^2/\nu$=1.50 (with a median of 1.48) 
for the flat X-ray slope sub-sample and $\chi^2/\nu$=2.00 (median 1.72) for
the step X-ray spectrum objects.

\subsection{\label{spec_var} Spectral variability}
\begin{figure}
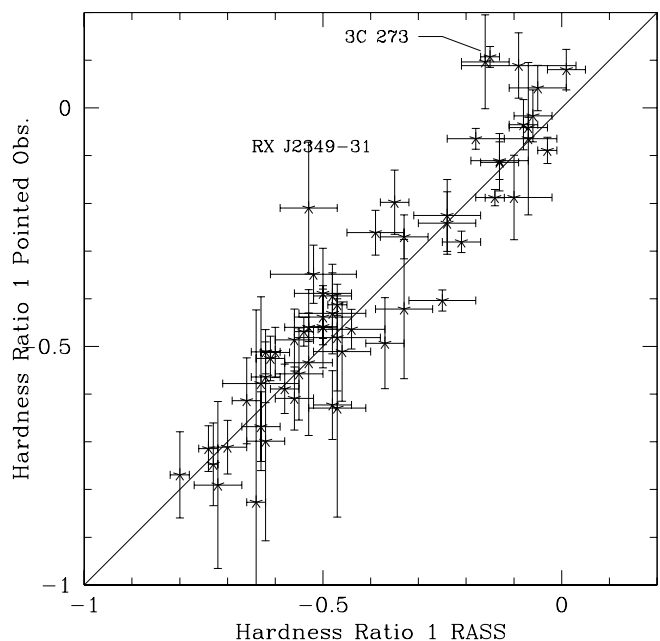

\parbox[h]{8.7cm}{
\clipfig{H2394_F5}{87}{06}{54}{200}{248}
}
\caption[ ] {\label{hr-hr} RASS vs. pointed observation hardness ratio}
\end{figure}

\begin{figure}
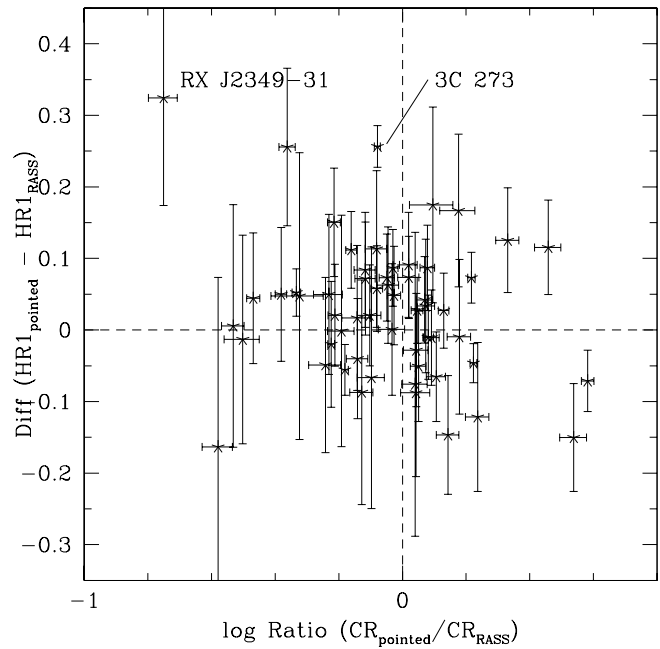

\parbox[h]{8.7cm}{
\clipfig{H2394_F6}{87}{06}{50}{200}{248}
}
\caption[ ] {\label{ratio_dif} Ratio of the count rates of 
pointed PSPC and RASS vs. the difference between their HR1. 
The transients IC 3599 and WPVS007 fall outside the boundaries of the plot.}
\end{figure}

Fig. \ref{hr-hr} shows the hardness ratios of the RASS vs. the pointed PSPC
observations. 
Most sources do not show dramatic spectral
changes. Only two objects, RX J2349.3--3126 (see
below) and 3C 273, may have varied.  
Indeed, the 
difference between the hardness
ratios of the RASS and the pointed PSPC observations is small: the mean 
hardness ratio increased by 0.03.
Fig. \ref{ratio_dif} displays the count rate 
ratio vs. the difference in the hardness ratios between the RASS and the later
pointed PSPC observations. The purpose of this plot is to 
demonstrate how the sources
have changed their spectra with count rate. Obviously, there is no clear correlation between both quantities, the changes going
in all possible directions. Spectral changes can also be studied, of course, by
using the spectral index \ax.   However, the \ax~ depend on model fitting.
We prefer to compare hardness
ratios because they are determined directly from the observed counts,
and the uncertainties are more simply interpreted. 
The mean change in hardness ratio for small count rate changes (ratios between 0.8 and 1.2) turns
out to be $0.04 \pm 0.01$. This might be due to some small losses of hard photons by the off-axis
correction for the RASS data as compared to the pointing data, which are usually taken on-axis,
and therefore not be an intrinsic property of our sample.

\subsection{\label{spec_ana} Spectral analysis}

Table \ref{spec_list} lists the spectral analysis of the RASS and pointed
PSPC data.  For the power-law fits to
the RASS and pointed PSPC data with the absorption parameter $N_H$ fixed to the
Galactic value, 
the X-ray fluxes are given
in the restframe ROSAT energy range between 0.2-2.0 keV. 

The mean X-ray slope of the RASS observation using a power-law fit with fixed 
$N_H$ is \ax=1.81 with a median of
1.67.  This is slightly flatter than  we found for our original sample
($<$\ax$>$=2.10; Grupe et al. 1998a). 
The mean X-ray
luminosity derived from this spectral model is log $L_X$=36.87 [W] with a 
median of 36.91.  The object with the highest luminosity
is 3C 273 (log $L_{\rm X}$=38.4). Dividing the sample into  low and high luminosity
subsamples using the median of log $L_X$ as the borderline, we find a difference in the
distribution of the X-ray slope \ax. The low-luminosity sub-sample has a mean of
$<$\ax$>=1.61$ (median=1.54), the high-luminosity sub-sample $<$\ax$>=1.89$ (median=1.92).

\subsection{\label{x-trans} Individual X-ray transient AGN}
Here, we present new results of HRI observations of the X-ray transient
AGN IC3599 and WPVS007. We briefly review the other most recently discovered
transients, RX J1624.9+7554 and RX J2217.9--5941. 

\subsubsection{IC 3599}
\begin{figure}
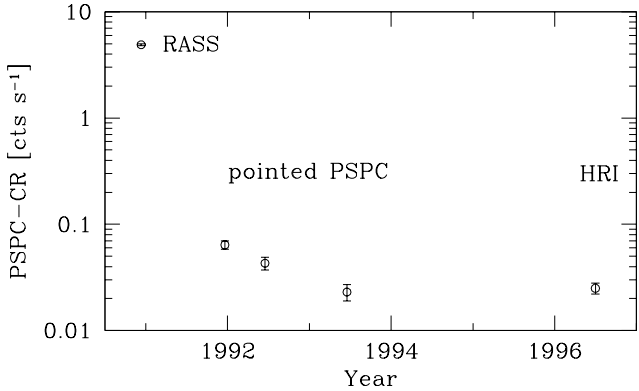

\parbox[h]{8.7cm}{
\clipfig{H2394_F7}{87}{10}{148}{162}{245}
}
\caption[ ] {\label{ic3599_cr} Long-term light curve of IC3599. The HRI count
rate was converted into a PSPC count rate (see Sect. \ref{obs}).}
\end{figure}

The Seyfert 2 galaxy
IC 3599 was one of the brightest AGN of our sample during the RASS
(4.90 PSPC cts s$^{-1}$) and was seen even with the Wide Field Camera (WFC) that
was attached to ROSAT (Pounds et al. 1993, Edelson et al. 1999).
It became fainter in pointed PSPC
observations in the following years by a factor of about 100
(see Grupe et al. 1995a, Brandt et al.
1995). 
We made two ROSAT HRI observations between 1996-06-30 and 1996-07-02
to monitor its long term behaviour. 
In order to get better
photon statistics on the source, both event files were merged into one. 
From this
merged event file with a total exposure time of 31621 s an HRI count rate of
4.62\pl0.49 $10^{-3}$ was measured, which corresponds to 0.025 PSPC \cts~
(see Table \ref{CR_list}) and an X-ray luminosity in the 0.2-2.0 keV band of log
$L_X$=33.6.
Fig. \ref{ic3599_cr} displays the long-term light curve.  
IC 3599 has shown a
response in its optical emission line spectrum. Shortly after the RASS, highly
ionized iron lines like FeXIV were observed (see Brandt et al. 1995). A year
later, these lines disappeared but other lower ionized iron lines like FeVII and
FeX appeared (Grupe et al. 1995a).

\begin{landscape}

{\small
\begin{table}
\caption{\label{spec_list} Spectral analysis of the RASS and pointed PSPC observations. The absorption
parameters $N_{\rm H,fit}$ and $N_{\rm H,gal}$ are given in units of $10^{20}~\rm cm~s^{-1}$ and the 
X-ray flux $F_X$  in $\rm Watts~m^{-2}$ in the rest-frame energy band between 0.2-2.0 keV. 
}
\begin{tabular}{rlccrccrccrcrl}
\hline\noalign{\smallskip}
 & & \multicolumn{7}{c}{RASS} & \multicolumn{3}{c}{PSPC-pointed}  \\
\rb{No.} & \rb{name}  & $N_{\rm H,fit}$   & \ax  & $\chi^2/\nu$ & $N_{\rm H. gal}$ & \ax  & $\chi^2/\nu$ & 
log F$_{\rm X}$ & \ax  & $\chi^2/\nu$ & log F$_{\rm X}$ & comments
\\
\noalign{\smallskip}\hline\noalign{\smallskip}
  1 & Mkn 335      &   4.49\pl1.36 & 2.29\pl0.29 & 15/19  &  3.96 & 2.10\pl0.10 & 16/20  & -13.40 & 2.13\pl0.01 & 233/178 & -13.47  \\
  2 & ESO 242-G8   &   0.70\pl1.14 & 1.13\pl0.66 & 15/10  &  1.48 & 1.56\pl0.26 & 17/11  & -14.31 & 1.64\pl0.05 & 53/56   & -14.45 \\
  3 & WPVS 007     &   0.68\pl0.71 & 3.10\pl0.99 & 5.4/13 &  2.82 & 8.70\pl1.75 & 8.9/14 & -14.27 & --- & --- & --- & 1 \\
  4 & RX J0057-22  &   1.37\pl0.51 & 1.82\pl0.27 & 25/25  &  1.48 & 1.89\pl0.10 & 25/26  & -13.77 & 1.99\pl0.03 & 78/93   & -13.63 \\
  5 & QSO 0056-36  &   1.73\pl1.19 & 1.62\pl0.51 & 29/22  &  1.94 & 1.72\pl0.16 & 29/23  & -14.29 & --- & --- & --- \\
  6 & RX J0100-51  &   2.49\pl1.59 & 1.75\pl0.52 & 16/18  &  2.42 & 1.73\pl0.18 & 16/19  & -13.99 & --- & --- & --- \\
  7 & RX J0105-14  &   2.34\pl0.91 & 1.53\pl0.30 & 30/19  &  1.77 & 1.29\pl0.09 & 32/20  & -13.89 & --- & --- & --- \\
  8 & RX J0117-38  &   2.82\pl1.35 & 2.41\pl0.44 & 3.9/9  &  2.08 & 2.09\pl0.15 & 5.4/10 & -14.18 & --- & --- & --- \\
  9 & MS 0117-28   &   2.13\pl1.56 & 2.51\pl0.66 & 6.8/13 &  1.65 & 2.27\pl0.26 & 7.2/14 & -14.54 & 2.70\pl0.12 & 28/31   & -14.46 \\
 10 & RX J0128-18  &   1.82\pl1.31 & 1.64\pl0.55 & 14/19  &  1.62 & 1.55\pl0.17 & 14/20  & -14.26 & --- & --- & --- \\
 11 & IRAS01267-21 &   1.51\pl1.26 & 1.55\pl0.62 & 14/15  &  1.28 & 1.43\pl0.19 & 14/16  & -14.29 & --- & --- & --- \\
 12 & RX J0148-27  &   2.54\pl0.82 & 2.62\pl0.30 & 21/22  &  1.50 & 2.12\pl0.11 & 30/23  & -13.90 & 1.88\pl0.03 & 178/110 & -13.90 \\
 13 & RX J0152-23  &   1.25\pl0.68 & 1.75\pl0.39 & 17/13  &  1.10 & 1.67\pl0.13 & 17/14  & -14.26 & --- & --- & --- \\
 14 & Mkn 1044     &   3.99\pl1.38 & 2.04\pl0.32 & 21/17  &  3.16 & 1.74\pl0.10 & 22/18  & -13.52 & 1.75\pl0.03 & 144/118 & -13.52 \\
 15 & Mkn 1048     &   3.18\pl1.62 & 1.67\pl0.43 & 13/9   &  2.83 & 1.53\pl0.14 & 13/10  & -13.84 & 1.51\pl0.03 & 181/140 & -13.77 \\
 16 & RX J0311-20  &   1.02\pl1.15 & 8.63\pl0.56 & 23/17  &  2.37 & 1.47\pl0.19 & 27/18  & -14.09 & --- & --- & --- \\
 17 & RX J0319-26  &   0.52\pl1.09 & 1.39\pl0.73 & 6.3/7  &  1.32 & 1.79\pl0.31 & 7.1/8  & -14.55 & 1.42\pl0.03 & 88/83   & -14.23 \\
 18 & RX J0323-49  &   2.45\pl0.85 & 2.35\pl0.29 & 36/21  &  1.72 & 2.03\pl0.10 & 39/22  & -14.02 & 1.83\pl0.05 & 79/55   & -14.39 \\
 19 & ESO 301-G13  &   1.66\pl0.91 & 1.75\pl0.41 & 20/11  &  2.19 & 2.01\pl0.14 & 21/12  & -14.09 & --- & --- & --- \\
 20 & VCV0331-37   &   1.48\pl1.76 & 1.51\pl0.89 & 6.0/7  &  1.63 & 1.59\pl0.27 & 6.0/8  & -14.32 & --- & --- & --- \\
 21 & RX J0349-47  &   0.78\pl0.80 & 2.04\pl0.58 & 13/12  &  1.44 & 2.45\pl0.33 & 15/13  & -14.75 & --- & --- & --- \\
 22 & Fair 1116    &   2.43\pl1.58 & 1.87\pl0.53 & 20/17  &  3.84 & 2.48\pl0.20 & 22/18  & -13.56 & --- & --- & --- \\
 23 & CTS 44       &   1.08\pl0.39 & 1.33\pl0.20 & 27/35  &  1.24 & 1.41\pl0.07 & 27/36  & -13.66 & --- & --- & --- \\
 24 & RX J0412-47  &   0.82\pl1.39 & 1.36\pl0.81 & 2.8/6  &  1.42 & 1.66\pl0.31 & 3.2/7  & -14.34 & --- & --- & --- \\
 25 & RX J0426-57  &   1.12\pl0.31 & 1.62\pl0.20 & 44/30  &  2.25 & 2.20\pl0.07 & 70/31  & -13.53 & 2.10\pl0.02 & 333/154 & -13.19 \\
 26 & Fair 303     &   1.69\pl1.12 & 1.89\pl0.51 & 13/20  &  0.99 & 1.51\pl0.17 & 15/21  & -14.19 & 1.31\pl0.05 & 58/53   & -14.24 \\
 27 & RX J0437-47  &   1.07\pl0.46 & 1.74\pl0.31 & 28/21  &  1.69 & 2.09\pl0.12 & 33/22  & -14.14 & 1.93\pl0.03 & 184/91  & -13.96 \\
 28 & RX J0439-53  &   1.35\pl1.85 & 2.28\pl1.04 & 3.1/5  &  1.55 & 2.39\pl0.42 & 32/6   & -14.44 & --- & --- & --- \\
 29 & H 0439-272   &   1.98\pl3.99 & 1.30\pl1.57 & 6.8/3  &  2.43 & 1.51\pl0.39 & 6.8/4  & -14.26 & 1.27\pl0.04 & 90/88   & -14.04 \\
 30 & 1 ES 0614-58 &   4.37\pl1.00 & 2.38\pl0.23 & 20/30  &  4.60 & 2.46\pl0.08 & 20/31  & -13.90 & --- & --- & --- \\
 31 & RX J0859+48  &   1.99\pl1.46 & 1.44\pl0.57 & 16/17  &  2.05 & 1.46\pl0.17 & 16/18  & -14.22 & --- & --- & --- \\
 32 & RX J0902-07  &   3.54\pl2.45 & 2.12\pl0.62 & 16/13  &  3.68 & 2.17\pl0.22 & 16/14  & -14.08 & --- & --- & --- \\
 33 & Mkn 110      &   1.47\pl0.64 & 1.25\pl0.30 & 13/21  &  1.56 & 1.29\pl0.09 & 13/22  & -13.86 & 1.47\pl0.01 & 290/182 & -13.26 \\
 34 & PG 0953+41   &   2.30\pl1.26 & 2.26\pl0.46 & 7.6/8  &  1.14 & 1.65\pl0.15 & 13/9   & -14.34 & 1.55\pl0.04 & 47/73   & -14.28 \\
 35 & RX J1005+43  &   1.65\pl0.96 & 2.15\pl0.46 & 9.4/9  &  1.08 & 1.81\pl0.16 & 11/10  & -14.49 & --- & --- & --- \\
 36 & RX J1007+22  &   6.20\pl2.79 & 2.91\pl0.54 & 15/21  &  2.69 & 1.68\pl0.16 & 24/22  & -14.19 & --- & --- & --- \\
 37 & CBS 126      &   1.36\pl0.62 & 1.62\pl0.33 & 18/19  &  1.41 & 1.65\pl0.10 & 18/20  & -14.03 & --- & --- & --- \\
 38 & HS 1019+37   &   1.32\pl0.95 & 9.83\pl0.43 & 12/9   &  1.28 & 0.98\pl0.12 & 12/10  & -14.23 & --- & --- & --- \\
 39 & Mkn 141      &   1.68\pl1.30 & 1.84\pl0.58 & 15/16  &  1.07 & 1.53\pl0.18 & 16/17  & -14.54 & --- & --- & --- \\
 40 & Mkn 142      &   1.61\pl0.55 & 2.10\pl0.27 & 34/26  &  1.18 & 1.88\pl0.10 & 36/27  & -14.04 & 1.77\pl0.03 & 139/85  & -14.20 \\
\noalign{\smallskip}\hline
\end{tabular}
\end{table}
}

\newpage
\clearpage

{\small
\begin{table}
\begin{tabular}{rlccrccrccrcrl}
\hline\noalign{\smallskip}
 & & \multicolumn{7}{c}{RASS} & \multicolumn{3}{c}{PSPC-pointed}  \\
\rb{No.} & \rb{name}  & $N_{\rm H,fit}$   & \ax  & $\chi^2/\nu$ & $N_{\rm H. gal}$ & \ax  & $\chi^2/\nu$ & 
log F$_{\rm X}$ & \ax  & $\chi^2/\nu$ & logF$_{\rm X}$ & comments \\
\noalign{\smallskip}\hline\noalign{\smallskip}
 41 & RX J1034+39  &   2.71\pl0.79 & 3.00\pl0.31 & 13/25  &  1.50 & 2.38\pl0.11 & 26/26  & -13.85 & 2.42\pl0.03 & 376/91  & -13.67 \\
 42 & RX J1117+65  &   1.95\pl1.10 & 2.50\pl0.49 & 6.9/9  &  0.91 & 1.89\pl0.17 & 12/10  & -14.62 & --- & --- & --- \\
 43 & PG 1115+407  &   1.42\pl1.27 & 1.81\pl0.66 & 13/13  &  1.93 & 2.05\pl0.24 & 13/14  & -14.43 & 1.98\pl0.06 & 49/44   & -14.53 \\
 44 & Ton 1388     &   1.14\pl0.91 & 1.61\pl0.57 & 11/7   &  1.22 & 1.65\pl0.17 & 11/8   & -14.29 & 1.62\pl0.02 & 193/149 & -14.12 \\
 45 & EXO 1128+69  &   1.12\pl0.40 & 1.49\pl0.22 & 42/34  &  1.34 & 1.60\pl0.07 & 42/35  & -13.97 & --- & --- & --- \\
 46 & B2 1128+31   &   2.61\pl2.08 & 1.59\pl0.66 & 12/12  &  2.22 & 1.43\pl0.20 & 12/13  & -14.38 & --- & --- & --- \\
 47 & SBS 1136+579 &   1.42\pl1.22 & 1.77\pl0.63 & 18/16  &  1.00 & 1.54\pl0.19 & 18/17  & -14.54 & 1.58\pl0.19 & 4.1/6   & -15.38 \\
 48 & Z 1136+3412  &   1.81\pl1.24 & 1.72\pl0.52 & 19/19  &  2.04 & 1.81\pl0.17 & 19/20  & -14.14 & 1.81\pl0.09 & 22/24   & -14.36 \\
 49 & Was 26	   &   2.75\pl0.97 & 1.69\pl0.29 & 22/20  &  2.10 & 1.43\pl0.09 & 24/21  & -13.82 & 1.43\pl0.03 & 121/92  & -13.98 \\
 50 & CASG 855     &   1.64\pl1.03 & 1.33\pl0.47 & 9.4/9  &  1.80 & 1.40\pl0.13 & 9.5/10 & -13.95 & --- & --- & --- \\
 51 & Mkn 1310     &   2.31\pl1.37 & 1.35\pl0.47 & 14/8   &  2.43 & 1.39\pl0.14 & 14/9   & -14.01 & 1.13\pl0.12 & 18/14   & -14.71 \\
 52 & NGC 4051     &   2.23\pl0.59 & 2.02\pl0.22 & 31/36  &  1.37 & 1.62\pl0.07 & 41/37  & -13.60 & 1.65\pl0.01 & 361/177 & -13.92 \\
 53 & GQ Com	   &   1.94\pl1.23 & 1.30\pl0.50 & 27/21  &  1.66 & 1.18\pl0.15 & 27/22  & -14.37 & 1.20\pl0.02 & 131/140 & -14.40 \\
 54 & RX J1209+32  &   4.34\pl4.52 & 3.18\pl1.18 & 2.7/5  &  0.00 & 0.86\pl0.27 & 19/6   & -14.73 & 1.09\pl0.09 & 36/20   & -14.98 \\
 55 & PG 1211+143  &   3.04\pl1.02 & 2.12\pl0.29 & 22/21  &  2.75 & 2.00\pl0.09 & 2.2/22 & -13.74 & 1.82\pl0.04 & 84/78   & -13.86 \\
 56 & Mkn 766	   &   3.40\pl0.64 & 1.77\pl0.16 & 19/29  &  1.69 & 1.11\pl0.05 & 57/30  & -13.31 & 1.28\pl0.01 & 474/171 & -13.48 \\
 57 & 3C 273	   &   1.58\pl0.34 & 1.21\pl0.16 & 56/42  &  1.79 & 1.30\pl0.05 & 57/43  & -13.23 & 1.00\pl0.01 & 234/200 & -13.18 \\
 58 & RX J1231+70  &   1.95\pl0.75 & 1.49\pl0.29 & 16/22  &  1.67 & 1.38\pl0.08 & 16/23  & -14.12 & --- & --- & --- \\
 59 & Mkn 771	   &   3.64\pl1.88 & 2.46\pl0.51 & 19/19  &  2.05 & 1.83\pl0.16 & 24/20  & -14.32 & 1.64\pl0.03 & 123/126 & -13.87 \\
 60 & CBS 150	   &   3.41\pl2.01 & 2.94\pl0.62 & 20/18  &  1.44 & 2.13\pl0.20 & 26/19  & -14.71 & 1.89\pl0.09 & 23/27   & -14.60 \\
 61 & MCG+08-23-06 &   1.36\pl1.09 & 1.36\pl0.53 & 19/19  &  1.40 & 1.38\pl0.16 & 19/20  & -14.37 & --- & --- & --- \\
 62 & IC 3599	   &   4.04\pl0.71 & 3.37\pl0.21 & 55/32  &  1.34 & 2.17\pl0.06 & 169/33 & -13.60 & 2.61\pl0.35 & 11/6    & -15.59 \\
 63 & NGC 4593     &   1.57\pl0.94 & 1.19\pl0.42 & 15/11  &  2.28 & 1.47\pl0.12 & 16/12  & -13.46 & 1.19\pl0.05 & 54/54   & -13.78 \\
 64 & IRAS 1239+33 &   2.74\pl1.38 & 2.02\pl0.42 & 19/11  &  1.35 & 1.37\pl0.12 & 25/12  & -14.24 & 1.28\pl0.03 & 195/98  & -14.43 \\
 65 & PG 1244+026  &   1.08\pl0.99 & 1.44\pl0.53 & 21/16  &  1.75 & 1.79\pl0.20 & 22/17  & -14.06 & 1.85\pl0.03 & 118/83  & -14.02 \\
 66 & RX J1304+02  &   1.50\pl1.13 & 2.26\pl0.59 & 18/18  &  1.74 & 2.38\pl0.25 & 18/19  & -14.44 & --- & --- & --- \\
 67 & PG 1307+085  &   2.75\pl1.85 & 1.87\pl0.55 & 8.8/17 &  2.05 & 1.58\pl0.17 & 8.9/18 & -14.32 & 1.58\pl0.04 & 86/80   & -14.24 \\
 68 & RX J1314+34  &   1.93\pl1.08 & 2.35\pl0.45 & 9.5/9  &  0.99 & 1.88\pl0.16 & 13/10  & -14.51 & 1.76\pl0.14 & 24/11   & -14.87 \\
 69 & RX J1319+52  &   2.04\pl1.12 & 2.06\pl0.44 & 4.5/9  &  1.16 & 1.60\pl0.14 & 8.0/10 & -14.41 & --- & --- & --- \\
 70 & PG 1322+659  &   1.32\pl0.77 & 1.74\pl0.42 & 5.5/10 &  2.01 & 2.07\pl0.15 & 7.5/11 & -14.30 & 1.81\pl0.04 & 64/73   & -14.27 \\
 71 & IRAS 1334+24 &   1.36\pl0.50 & 2.01\pl0.27 & 20/25  &  1.12 & 1.88\pl0.10 & 21/26  & -13.92 & 1.83\pl0.05 & 57/53   & -14.20 \\
 72 & Ton 730	   &   1.29\pl0.99 & 1.96\pl0.57 & 13/16  &  1.05 & 1.83\pl0.20 & 13/17  & -14.59 & --- & --- & --- \\
 73 & RX J1355+56  &   1.85\pl0.92 & 2.31\pl0.42 & 15/10  &  1.15 & 1.93\pl0.15 & 18/11  & -14.47 & --- & --- & --- \\
 74 & PG 1402+25   &   1.52\pl1.01 & 1.85\pl0.50 & 4.0/7  &  1.48 & 1.83\pl0.17 & 4.0/8  & -14.42 & 1.91\pl0.06 & 51/49   & -14.12 \\
 75 & RX J1413+70  &   1.15\pl0.98 & 1.07\pl0.47 & 13/23  &  1.93 & 1.40\pl0.15 & 14/24  & -14.20 & --- & --- & --- \\
 76 & NGC 5548     &   1.47\pl0.38 & 1.13\pl0.17 & 47/30  &  1.93 & 1.33\pl0.05 & 52/31  & -13.35 & 1.50\pl0.01 & 324/201 & -13.10 \\
 77 & Q 1421-0013  &   5.99\pl3.21 & 2.71\pl0.62 & 17/14  &  2.97 & 1.72\pl0.19 & 22/15  & -14.23 & --- & --- & --- \\
 78 & Mkn 813	   &   1.29\pl0.87 & 1.11\pl0.42 & 8.2/10 &  2.54 & 1.64\pl0.14 & 13/11  & -14.12 & --- & --- & --- \\
 79 & Mkn 684	   &   1.68\pl1.38 & 1.44\pl0.61 & 12/15  &  1.50 & 1.36\pl0.18 & 13/16  & -14.40 & --- & --- & --- \\
 80 & Mkn 478	   &   1.27\pl0.26 & 2.22\pl0.16 & 20/32  &  1.04 & 2.08\pl0.06 & 24/33  & -13.64 & 2.18\pl0.05 & 68/54   & -13.95 \\
\noalign{\smallskip}\hline
\end{tabular}
\end{table}
}

\newpage
\clearpage

{\small
\begin{table}
\begin{tabular}{rlccrccrccrcrl}
\hline\noalign{\smallskip}
 & & \multicolumn{7}{c}{RASS} & \multicolumn{3}{c}{PSPC-pointed}  \\
\rb{No.} & \rb{name}  & $N_{\rm H,fit}$   & \ax  & $\chi^2/\nu$ & $N_{\rm H. gal}$ & \ax  & $\chi^2/\nu$ & 
log F$_{\rm X}$ & \ax  & $\chi^2/\nu$ & logF$_{\rm X}$ & comments \\
\noalign{\smallskip}\hline\noalign{\smallskip}
 81 & PG 1448+273  &   4.52\pl1.70 & 2.13\pl0.37 & 6.6/11 &  2.71 & 1.52\pl0.12 & 13/12  & -14.05 & --- & --- & --- \\
 82 & Mkn 841	   &   2.48\pl1.60 & 1.61\pl0.52 & 23/19  &  2.24 & 1.50\pl0.17 & 23/20  & -14.13 & 1.42\pl0.01 & 220/190 & -13.61 \\
 83 & HS 1529+56   &   1.77\pl0.73 & 1.78\pl0.32 & 20/18  &  1.15 & 1.46\pl0.10 & 24/19  & -14.32 & --- & --- & --- \\
 84 & Mkn 493	   &   1.51\pl1.33 & 1.26\pl0.62 & 9.0/16 &  2.22 & 1.57\pl0.18 & 9.7/17 & -14.28 & 1.60\pl0.04 & 55/66   & -14.39 \\
 85 & Mkn 876	   &   2.68\pl0.79 & 1.57\pl0.24 & 28/30  &  2.95 & 1.67\pl0.08 & 28/31  & -14.03 & 1.62\pl0.03 & 143/124 & -13.86 \\
 86 & RX J1618+36  &   1.57\pl0.67 & 1.69\pl0.31 & 15/19  &  1.26 & 1.54\pl0.10 & 15/20  & -14.24 & --- & --- & --- \\
 87 & KUG 1618+40  &   1.63\pl0.97 & 1.87\pl0.45 & 14/10  &  0.96 & 1.52\pl0.14 & 16/11  & -14.50 & --- & --- & --- \\
 88 & RX J1624+75  &   5.65\pl1.77 & 2.89\pl0.37 & 14/15  &  3.87 & 2.28\pl0.12 & 19/16  & -14.14 & --- & --- & --- \\
 89 & PG 1626+554  &   1.23\pl0.76 & 1.42\pl0.38 & 17/12  &  1.99 & 1.79\pl0.13 & 20/13  & -14.33 & 2.02\pl0.06 & 71/50   & -14.00 \\
 90 & EXO 1627+401 &   0.71\pl0.38 & 2.15\pl0.37 & 7.0/15 &  0.84 & 2.25\pl0.16 & 7.4/16 & -14.65 & 2.16\pl0.05 & 75/53   & -14.31 \\
 91 & HS 1702+32   &   2.12\pl1.03 & 1.99\pl0.39 & 9.1/11 &  2.42 & 2.13\pl0.14 & 9.4/12 & -14.31 & 2.43\pl0.05 & 44/57   & -14.27 \\
 92 & II Zw 136    &   2.53\pl1.10 & 1.37\pl0.34 & 16/15  &  4.65 & 2.10\pl0.12 & 26/16  & -13.60 & 2.08\pl0.03 & 181/145 & -13.57 \\
 93 & RX J2146-30  &   2.49\pl1.60 & 1.66\pl0.51 & 22/21  &  2.31 & 1.59\pl0.16 & 22/22  & -14.15 & --- & --- & --- \\
 94 & A 09.25	   &   2.03\pl1.31 & 1.63\pl0.50 & 8.2/20 &  1.47 & 1.37\pl0.15 & 9.0/21 & -14.14 & --- & --- & --- \\
 95 & NGC 7214     &   0.95\pl1.24 & 1.02\pl0.63 & 16/14  &  1.62 & 1.34\pl0.20 & 16/15  & -14.21 & 1.68\pl0.01 & 291/181 & -13.65 \\
 96 & RX J2216-44  &   1.23\pl0.64 & 1.98\pl0.38 & 1.1/12 &  2.17 & 2.48\pl0.17 & 16/13  & -14.02 & --- & --- & --- \\
 97 & RX J2217-59  &   1.36\pl0.84 & 2.08\pl0.46 & 7.3/8  &  2.58 & 2.69\pl0.21 & 12/9   & -14.18 & --- & --- & --- \\
 98 & PKS 2227-399 &   0.63\pl1.03 & 0.72\pl0.55 & 8.2/13 &  1.20 & 1.00\pl0.19 & 9.1/14 & -14.50 & 1.15\pl0.05 & 48/56   & -14.31 \\
 99 & RX J2242-38  &   2.69\pl1.81 & 2.92\pl0.70 & 16/16  &  1.18 & 2.19\pl0.26 & 2.0/17 & -14.64 & --- & --- & --- \\
100 & RX J2245-46  &   0.61\pl0.49 & 1.71\pl0.35 & 21/10  &  1.95 & 2.55\pl0.21 & 33/11  & -14.09 & 2.59\pl0.04 & 348/82  & -13.75 \\
101 & RX J2248-51  &   0.88\pl0.39 & 1.72\pl0.24 & 26/24  &  1.27 & 1.95\pl0.10 & 28/25  & -13.88 & 1.76\pl0.03 & 160/112 & -13.77 \\
102 & MS 2254-36   &   1.27\pl0.75 & 1.84\pl0.43 & 30/11  &  1.15 & 1.78\pl0.15 & 30/12  & -13.97 & 1.83\pl0.04 & 68/68   & -14.16 \\
103 & RX J2258-26  &   3.89\pl4.23 & 2.11\pl0.98 & 6.1/5  &  2.11 & 1.50\pl0.31 & 6.6/6  & -14.25 & --- & --- & --- \\
104 & RX J2301-59  &   1.40\pl1.04 & 1.10\pl0.48 & 7.0/8  &  2.69 & 1.65\pl0.15 & 11/9   & -14.00 & --- & --- & --- \\
105 & RX J2301-55  &   1.50\pl0.88 & 2.04\pl0.45 & 9.6/9  &  1.54 & 2.09\pl0.17 & 9.6/10 & -14.30 & 2.22\pl0.04 & 101/64  & -14.32 \\
106 & RX J2304-35  &   1.43\pl2.01 & 1.63\pl1.01 & 4.7/7  &  1.47 & 1.65\pl0.29 & 4.7/8  & -14.39 & 1.73\pl0.06 & 35/40   & -14.86 \\
107 & RX J2312-34  &   0.71\pl2.14 & 0.78\pl1.13 & 3.5/3  &  1.80 & 1.34\pl0.39 & 4.3/4  & -14.50 \\
108 & RX J2317-44  &   1.30\pl1.23 & 2.50\pl0.80 & 10/10  &  1.89 & 2.87\pl0.44 & 11/11  & -14.41 & --- & --- & --- \\
109 & RX J2325-32  &   1.50\pl2.90 & 2.01\pl1.44 & 0.8/2  &  1.33 & 1.92\pl0.49 & 0.8/3  & -14.57 & --- & --- & --- \\
110 & IRAS 23226-3 &   1.99\pl2.07 & 1.38\pl0.80 & 8.0/9  &  1.59 & 1.20\pl0.22 & 8.2/10 & -13.98 & --- & --- & --- \\
111 & MS 23409-151 &   3.65\pl2.29 & 2.60\pl0.63 & 16/14  &  2.20 & 2.03\pl0.21 & 18/15  & -14.23 & 2.09\pl0.02 & 139/118 & -14.03 \\
112 & RX J2349-31  &   1.02\pl1.01 & 1.55\pl0.57 & 18/14  &  1.23 & 1.67\pl0.22 & 18/15  & -14.50 & 1.17\pl0.07 & 26/25   & -15.06 \\
113 & AM 2354-304  &   1.55\pl1.41 & 1.39\pl0.68 & 9.5/14 &  1.37 & 1.30\pl0.19 & 9.6/15 & -14.39 & --- & --- & --- \\
\noalign{\smallskip}\hline
\end{tabular}
\end{table}
}

\end{landscape}

Monitoring this object in the optical until now shows that these
lines are slowly fading. 
The interpretation of the X-ray outburst in IC 3599 is that it is the result of
 disk instabilities or  tidal disruption of a
star by the central back hole (see Brandt et al. 1995 and Grupe et al. 1995a). A
similar outburst event has been reported in the galaxy NGC 5905 
(Komossa \& Bade 1999).

\subsubsection{WPVS007}
\begin{figure}
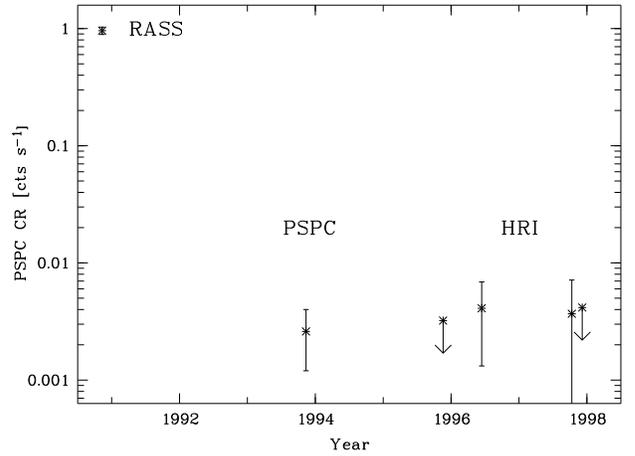

\parbox[h]{8.7cm}{
\clipfig{H2394_F8}{87}{10}{10}{275}{195}
}
\caption[ ] {\label{wpvs007_cr} Long-term light curve of WPVS007. The downwards
arrows mark the upper limits of the HRI observations converted into PSPC counts
(see Sect. \ref{obs}).}
\end{figure}

The Narrow-line Seyfert 1 galaxy
WPVS007 was the softest AGN detected during the RASS (Grupe et al.
1995b) and it was bright (about 1 \cts). Like
IC 3599 it was one of a few AGN that
were seen during the WFC all-sky survey (Edelson et al. 1999). 
After the RASS it was observed three years later by the  ROSAT PSPC
and was practically `turned-off'. In the following years we monitored the source
four times (see Table \ref{wpvs007_hri}) using the ROSAT HRI. In two of these
observations WPVS007 was detected again (ROR 702705 and 702923). For the
other two only upper limits can be given. 
Fig. \ref{wpvs007_cr} shows the new long-term light curve of WPVS007.  
The HRI count rates of WPVS007 convert to about log $L_X$ =31
[W].
Like IC 3599 we monitored
WPVS007's optical spectrum over several years. So far we not detected any
change in the optical spectrum.
A possible interpretation of
the transience in WPVS007 is that the temperature of the Comptonization layer
above the disk that scatters the thermal UV disk photons into the soft X-ray
range, changed. A lower temperature caused a shift of the EUV bump spectrum out
of the ROSAT energy range of 0.1-2.4 keV. This explains why the bolometric
luminosity changed by a factor of 2 but the PSPC count rate decreased by a factor
of 400 (Grupe et al. 1995b). It is also interesting to note that WPVS007 was
 detected by the HRI considering that  no hard photons
above 0.5 keV were detected in the RASS observation\footnote{The 
HRI has a lower efficiency in soft X-rays than the PSPC}.

\begin{table}
\begin{flushleft}
\caption{\label{wpvs007_hri} Pointed HRI observations of WPVS007. The UT
observing
dates are given as yymmdd, the exposure time is in s, 
the HRI count rate in
units of $\rm 10^{-4}~HRI~cts~s^{-1}$, and the converted PSPC count rate in
units of $\rm 10^{-3}~PSPC~cts~s^{-1}$. $3\sigma$ 
Upper limits are marked as (u.l.). 
}
\begin{tabular}{lccrcc}
\hline\noalign{\smallskip}
\# & ROR & obs. date & $\rm T_{obs}$ & HRI CR & PSPC CR \\
\noalign{\smallskip}\hline
1 & 702703 & 951118 & 15994 & 4.04 (u.l.) & 3.22 \\
2 & 702705 & 960613 & 19863 & 5.14\pl3.48 & 4.10 \\
3 & 702923 & 971110 & 10851 & 4.61\pl4.33 & 3.68 \\
4 & 702921 & 971206 &  9976 & 5.20 (u.l)  & 4.16 \\
& all   &       --- & 56684 & 5.29\pl2.00 & 4.22 \\
\noalign{\smallskip}\hline
\end{tabular}
\end{flushleft}
\end{table}

\subsubsection{RX J1624.9+7554}
This X-ray transient was discovered in a non-active spiral galaxy 
(Grupe et al. 1999b). 
Between the RASS and the pointed PSPC observations years later
the object vanished completely
in X-rays. Because this source is in its normal state a non-active galaxy, it is
most likely that its X-ray outburst
 was caused by a tidal disruption of a star by the central black hole.
Komossa and Greiner (1999) and most recently Greiner et al. (2000)
also reported about other 
X-ray outbursts in a non-active galaxies.

\subsubsection{RX J2217.9--5941}

This Narrow-line Seyfert 1 galaxy has shown a decrease in its count rate by a
factor of about 30 between the RASS and its last HRI observation in 1998
(Grupe et al. 2000, in prep.).
It is not
yet clear if this is a real transient source, which means the the count rate is
still decreasing unti today, 
or if it was observed
 in its `low-state' in
both HRI observations in 1997 and 1998. 
The RASS light curve
shows a decline in count rate by a factor of approximately 15 in about two
days. However, the HRI count rates do not show strong variability over their
four day coverage.

\subsubsection{RX J2349.3--3126}

RX J2349.3--3126 is the only source in the current sample that has shown a
significant change in its X-ray spectrum between the RASS and its pointed PSPC
observation two and a half years later. Its hardness ratio has 
changed from --0.53\pl0.06
during the RASS to --0.21\pl0.13 in the pointed observation (Table \ref{CR_list}).
It also showed a decrease in its mean count rate between the RASS and the 
pointed observation by a factor of about 6. 
RX J2349.3--3126 seems to show
large amplitude variations only over long time scales. During the RASS coverage the
source doubled its count rate in about two days. During the pointed PSPC observation
the source remained almost constant over about 6 hours.

Optically, RX J2349.3--3126 is a Seyfert 1 galaxy with very broad emission
lines (FWHM(H$\beta$)=7715\kms; Grupe et al. 1999a). An R-image taken with the ESO/MPG 2.2m
telescope at La Silla besides the very bright nucleus displays several bright spots, and tidal
tails. So this AGN may be fueled by a merger.

\section{\label{discuss} Discussion}

\subsection{X-ray transient AGN}
Why are there no `turn-ons' among the sources of our sample? 
This is a selection effect due to the definition of our sample. 
Because our sample is chosen to include only sources that were bright during
the RASS, we will be biased against sources that were faint then but have since
brightened, but we include sources that have since become fainter -- sometimes
much fainter -- than our original count rate limit. For obvious reasons,
only the brightest sources were re-observed in later pointed observations. 
Fainter RASS
sources can be seen only serendipitously brighter in pointed observations.
Because of the much smaller coverage of the sky in the pointed observations
(15\% PSPC, 2\% HRI; Voges, private communications)
there is not much chance to detect a transient in its `high' state. 
So far only one
AGN has been reported being faint in the RASS and bright in a later pointed
PSPC observation, RX J1242.6--1119 (Komossa \& Greiner 1999).

{\it What causes transience in AGN?} 
 The most common explanation of X-ray transience in AGN is a sudden increase of
the accretion rate. This could be the
result of either accretion disk instabilities, or even the tidal disruption of
a star as has been suggested by Rees (1990). While in the Seyfert 2
galaxy IC 3599 both accretion disk instabilities or a tidal disruption of a
star might cause the X-ray outburst (Brandt et al. 1995, Grupe et al. 1995a)
in  non active galaxies like
RX J1624.9+7554, RX J1242.6--1119, NGC 5905 (Komossa \& Bade 1999) or 
RX J1420.4+5334 (Greiner et al. 2000) tidal
disruption seems to be the more likely explanation. 
In the case of the X-ray transient WPVS007,  Grupe et al. (1995b) 
discussed the possibility of a temperature change in the Comptonization layer
above the disk as an alternative explanation. 
A lower temperature of this layer would shift the soft X-ray
spectrum out of the ROSAT PSPC energy window (0.1-2.4 keV) and would mimic a
dramatic change in the X-ray/Big Blue Bump flux.

Converting the HRI count rates of the two transients IC 3599 and WPVS007 into
X-ray fluxes shows that they now have typical X-ray luminosities of a normal
galaxy (about log $L_X$=33 [W]). In both cases, we can consider this as their
pre-outburst/transient luminosities. 

\subsection{X-ray variability}

We find that low-luminosity objects have a higher probability of being
found to be variable than the high-luminosity ones
(Fig. \ref{var_check}).
 This result is not as prominent as in the samples
of Boller et al. (1996) and Leighly (1999a,b). The reason is that their
samples stretch over 7 and 5 orders in luminosity while ours only stretches
over $\approx$2 orders. However, in our Principal Component Analysis on our
original sample (Grupe et al. 1998a, 1999a) we found this result as a part of
the Eigenvector 1 relationship. 
There are two explanations for the dependence of the
variability on luminosity: a) the size of the Black Hole/AGN engine, and
b) the number of the X-ray emitting regions (see discussion in Leighly
1999a).
The $\chi^2/\nu$ test for the
short-term variability in the RASS data is a robust test. It does not take the
length of the observation into account. However, the RASS coverages are
usually in the order of days and therefore comparable. Only for a few sources
the RASS observations were split into two parts half a year apart. 
 We should  mention that the result of the distribution of the variability
seen in Fig. \ref{var_check}
is smeared out
when the excess variance (e.g. Nandra et al. 1997, Leighly 1999a)
is used instead if $\chi^2/\nu$.

The comparison of the long-term variability (right panel of Fig
\ref{var_check}) is more complicated. The 
time gaps between RASS and pointed observation can
vary from source to source between half a year up to 8 years. 
This is the reason why the HRI data points
suggest less variability than the PSPC data, because they have been observed
later than the PSPC. 
The only way to check out the long term variability of a large sample of AGN is
repeated monitoring of the sky like it is performed for example by RXTE's
All-Sky Monitor. 

Fig. 3 in Leighly 1999a shows that NLS1 are more variable than Broad-line
Seyfert 1s. In our sample a large percentage are NLS1 (Grupe et al. 1999a,
Grupe et al. 2001 in prep). NLS1 are also objects that have the steepest
X-ray spectra (Boller et al. 1996, Grupe et al. 1998a). Therefore we
checked for a relation between  \ax~ and
the strength of the variability. This is what we find for the short-term
variability throughout the sample: objects with steeper X-ray spectra
(preferentially NLS1) show stronger variability than those with flatter X-ray
spectra (see Fig. \ref{ax_chisq}). This is in agreement with the 
findings of Green et al. (1993) where `sources with steeper energy spectra have
higher normalized variability amplitudes'.

In principle, the same intrinsic processes that apply to X-ray
transience (changes in the accretion rate or the disk temperature) also apply to
the normal variability, but on a much lower level. In cases of very rapid
variability, such as found in IRAS13224--3809 (Boller et al. 1997), 
relativistic and Doppler boosting
and gravitational lensing effects (see Boller et al. 1997, Leighly 1999a) have
to be taken into account. 
The variability will be stronger amplified in steep X-ray sources.

Another alternative explanation of variability is a change in the cold and warm
absorber column densities
and their ionization states (e.g. Abrassart \& Czerny 2000,
Komossa \& Meerschweinchen 2000).

We have shown that ROSAT with its All-Sky Survey and the later pointed
observations was a well suited
experiment to detect X-ray transient sources and to
monitor the long-term behaviour of AGN. The best way to find more transients 
would be to perform all-sky surveys repeatedly.

\acknowledgements{We thank Drs Bev Wills, Mario Gliozzi,
Wolfgang Voges, Stefanie Komossa, and Joachim Tr\"umper
 for useful suggestions and
discussions. We also want to thank our referee, Prof. Dr. A. Lawrence for his
comments on the manuscript and valuable information on additional 
references.
 This research has made use of the NASA/IPAC
Extragalactic Database (NED) which is operated by the Jet Propulsion
Laboratory, Caltech, under contract with the National Aeronautics and
Space Administration.  The ROSAT project is
supported by the Bundesministerium f\"ur Bildung, Wissenschaft,
Forschung und Technologie (BMBF) and the Max-Planck-Gesellschaft.

This paper can be retrieved via WWW:
http://www.xray.mpe.mpg.de/~dgrupe/research/refereed.html}

\end{document}

%% file: clipfig.tex
\def\clipfig#1{\def\lbracket{[}\def\testit{#1}%
    \ifx\testit\lbracket\let\next=\optclipfig\else\let\next=\stdclipfig\fi%
    \next{#1}}
%
\newcommand {\hclipfig} [7] {\clipfig[#7]{#1}{#2}{#3}{#4}{#5}{#6}}
%
\def\usemodepsfig {\global\def\cfmode{x}\typeout{*** set clipfig to PSFIG mode ***}}
\def\usemodeepsf  {\global\def\cfmode{}\typeout{*** set clipfig to EPSF mode ***}}
\def\useunitmm    {\global\def\cfunit{x}\typeout{*** set clipfig to use mm as unit ***}}
\def\useunitcm    {\global\def\cfunit{}\typeout{*** set clipfig to use cm as unit ***}}
\def\clipfigsettings {\ifx\cfmode\empty\def\ccfmode{EPSF }\else\def\ccfmode{PSFIG }\fi%
    \ifx\cfunit\empty\def\ccfunit{cm }\else\def\ccfunit{mm }\fi%
    \typeout{*** current clipfig settings: \ccfmode mode, using \ccfunit as unit ***}}
%
%
%
%
\def\stdclipfig#1#2#3#4#5#6{\ifx\cfmode\empty%
    \let\next=\eclipfig\else\let\next=\pclipfig\fi%
    \next{#1}{#2}{#3}{#4}{#5}{#6}}
\def\optclipfig#1#2]#3#4#5#6#7#8{\ifx\cfmode\empty%
    \let\next=\ehclipfig\else\let\next=\phclipfig\fi%
    \next{#3}{#4}{#5}{#6}{#7}{#8}{#2}}
%
%
%
\newcommand {\pclipfig}[6] {\ifx\cfunit\empty%
        \psfig{figure=#1.ps,width=#2cm,bbllx=#3cm,bblly=#4cm,bburx=#5cm,%
           bbury=#6cm,clip=}\else%
        \psfig{figure=#1.ps,width=#2mm,bbllx=#3mm,bblly=#4mm,bburx=#5mm,%
           bbury=#6mm,clip=}\fi}
\newcommand {\phclipfig}[7] {\ifx\cfunit\empty%
        \hspace{#7cm}\psfig{figure=#1.ps,width=#2cm,bbllx=#3cm,bblly=#4cm,%
           bburx=#5cm,bbury=#6cm,clip=}\else%
        \hspace{#7mm}\psfig{figure=#1.ps,width=#2mm,bbllx=#3mm,bblly=#4mm,%
           bburx=#5mm,bbury=#6mm,clip=}\fi}
%
%
%
\newcommand {\eclipfig}[6]{%
  \ifx\cfunit\empty\epsfxsize=#2cm\else\epsfxsize=#2mm\fi%
  \epsfclipon\epsfverbosetrue%
  \cfcmtopspts{#3}\cfllxi=\cftempi\cfllxf=\cftempf%
  \cfcmtopspts{#4}\cfllyi=\cftempi\cfllyf=\cftempf%
  \cfcmtopspts{#5}\cfurxi=\cftempi\cfurxf=\cftempf%
  \cfcmtopspts{#6}\cfuryi=\cftempi\cfuryf=\cftempf%
  \def\cfstra{\number\cfllxi.\number\cfllxf}%
  \def\cfstrb{\number\cfllyi.\number\cfllyf}%
  \def\cfstrc{\number\cfurxi.\number\cfurxf}%
  \def\cfstrd{\number\cfuryi.\number\cfuryf}%
  \hbox{\epsfbox[{\cfstra} {\cfstrb} {\cfstrc} {\cfstrd}]{#1.ps}}}
\newcommand {\ehclipfig}[7]{%
  \ifx\cfunit\empty\epsfxsize=#2cm\else\epsfxsize=#2mm\fi%
  \epsfclipon\epsfverbosetrue%
  \cfcmtopspts{#3}\cfllxi=\cftempi\cfllxf=\cftempf%
  \cfcmtopspts{#4}\cfllyi=\cftempi\cfllyf=\cftempf%
  \cfcmtopspts{#5}\cfurxi=\cftempi\cfurxf=\cftempf%
  \cfcmtopspts{#6}\cfuryi=\cftempi\cfuryf=\cftempf%
  \def\cfstra{\number\cfllxi.\number\cfllxf}%
  \def\cfstrb{\number\cfllyi.\number\cfllyf}%
  \def\cfstrc{\number\cfurxi.\number\cfurxf}%
  \def\cfstrd{\number\cfuryi.\number\cfuryf}%
  \ifx\cfunit\empty\hspace{#7cm}\else\hspace{#7mm}\fi%
  \hbox{\epsfbox[{\cfstra} {\cfstrb} {\cfstrc} {\cfstrd}]{#1.ps}}%
  \vspace{-1mm}}
%
%
%
\newdimen\cfllxi \newdimen\cfllyi  \newdimen\cfurxi  \newdimen\cfuryi
\newdimen\cfllxf \newdimen\cfllyf  \newdimen\cfurxf  \newdimen\cfuryf
\newdimen\cftemp \newdimen\cftempi \newdimen\cftempf
\newdimen\cfpspoint \cfpspoint=1bp
%
%
%
\newcommand{\cfcmtopspts}[1]{\ifx\cfunit\empty%
  \cftemp=#1cm\else\cftemp=#1mm\fi%
  \multiply\cftemp10\divide\cftemp\cfpspoint%
  \cftempf=\cftemp\divide\cftemp10\cftempi=\cftemp\multiply\cftemp10%
  \advance\cftempf-\cftemp}
%
%
\def\cfmode{}\def\cfunit{}\clipfigsettings
%